\documentclass[sigconf]{acmart}

\renewcommand\footnotetextcopyrightpermission[1]{} % removes footnote with conference information in first column
\usepackage{gensymb}
\usepackage{booktabs} % For formal tables
\usepackage{multirow}
\usepackage{subcaption}

\begin{document}
\title{WarpFlow: Exploring Petabytes of Space-Time Data}

\author{Catalin Popescu}
\affiliation{%
  \institution{Google, Inc.}
  % \streetaddress{1600 Amphitheatre Parkway}
  % \city{Mountain View}
  % \state{California}
  % \postcode{94043}
}
\email{cpopescu@google.com}

\author{Deepak Merugu}
\affiliation{%
  \institution{Google, Inc.}
  % \streetaddress{1600 Amphitheatre Parkway}
  % \city{Mountain View}
  % \state{California}
  % \postcode{94043}
}
\email{deepakmerugu@gmail.com}

\author{Giao Nguyen}
\affiliation{%
  \institution{Google, Inc.}
  % \streetaddress{1600 Amphitheatre Parkway}
  % \city{Mountain View}
  % \state{California}
  % \postcode{94043}
}
\email{giao@google.com}

\author{Shiva Shivakumar}
\affiliation{%
  \institution{Google, Inc.}
  % \streetaddress{1600 Amphitheatre Parkway}
  % \city{Mountain View}
  % \state{California}
  % \postcode{94043}
}
\email{shiva@google.com}

\renewcommand{\shortauthors}{C. Popescu et al.}

\begin{abstract}

WarpFlow is a fast, interactive data querying and processing system with a
focus on petabyte-scale spatiotemporal datasets and \emph{Tesseract} queries.
With the rapid growth in smartphones and mobile navigation services, we now
have an opportunity to radically improve urban mobility and reduce friction in
how people and packages move globally every \emph{minute-mile}, with data.
WarpFlow speeds up three key metrics for data engineers working on such
datasets -- \emph{time-to-first-result}, \emph{time-to-full-scale-result}, and
\emph{time-to-trained-model} for machine learning.

\end{abstract}



% \keywords{ACM proceedings, \LaTeX, text tagging}

\maketitle

\section{Introduction}

Analytical data processing plays a central role in product development, and
designing and validating new products and features. Over the past few years, we
have seen a surge in demand for petascale interactive analytical query engines
(e.g., Dremel \cite{melnik:dremel}, F1 \cite{shute:f1}, Shasta
\cite{manoharan:shasta}), where developers execute a series of SQL queries over
datasets for iterative data exploration. Also, we have seen a tremendous growth
in petascale batch pipeline systems (e.g., MapReduce \cite{dean:mr}, Flume
\cite{chambers:flume}, Spark \cite{zaharia:spark}), where developers express
map-reduce and parallel-do style processing over datasets in batch mode.

In this paper, we focus on an important query pattern of ad hoc
\emph{Tesseract}\footnote{In geometry, Tesseract is the four-dimensional analog
of a cube. It is popularized as a spatiotemporal hyper-cube in the film
\emph{Interstellar}, and as the cosmic cube containing the Space Stone with
unlimited energy in the \emph{Avengers} film series.} queries. These are``big
multi-dimensional joins'' on spatiotemporal datasets, such as datasets from
Google Maps, and the fast-growing passenger ride-sharing, package delivery and
logistics services (e.g., Uber, Lyft, Didi, Grab, FedEx). For example, a
billion people around the world use Google Maps for its navigation and traffic
speed services \cite{misc:gmapsusers}, finding their way along 1 billion
kilometers each day \cite{misc:gmapskm}. To constantly improve the service on
every \emph{minute-mile}, engineers answer questions such as: (1) which roads
have high speed variability and how many drivers are affected, (2) how many
commuters, in aggregate, have many travel modes, e.g., bike after taking public
transit? (3) what are restaurant wait times when they are busy?

For such queries, we need to address a few important challenges:

\begin{itemize}

  \item How to analyze large, and often noisy spatiotemporal datasets? Most of
  this data comes from billions of smartphones moving through urban
  environments. These devices compute current location estimate by fusing GPS,
  WiFi, cellular signal and other available sensors, with varying degrees of
  accuracy (often 3 -- 30 meters away from the true position), based on urban
  canyon effects, weather and indoor obstructions \cite{zandbergen:gps,
  lee:gps, vonWatzdorf:gps}. These location observations are then recorded on
  the device and pushed to the cloud with an accuracy estimate. Also, each
  navigation path is recorded as a time-series of such noisy observations. A
  few key questions include: (a) how to store and index such rich data (e.g.,
  locations and navigation paths), (b) how to address noise with filtering and
  indexing techniques?

  \item How to speedup developer workflow while iterating on such queries? Each
  typical developer workflow begins with a new idea. The developer then tests
  the idea by querying the datasets, usually with simplified queries on small
  samples of the data. If the idea shows promise, they validate it with a
  full-scale query on all the available data. Depending on the outcome, they
  may repeat these steps several times to refine or discard the original idea.
  Finally, the developer pushes the refined idea towards production.

  One hurdle in this development cycle is the lengthy iteration time -- long
  cycles (several hours to days) prevent a lot of potential ideas from being
  tested and refined. This friction arises from: \emph{(i)} long
  \emph{commit-build-deploy} cycles when using compiled pipelines, and
  \emph{(ii)} composing complex queries on deeply nested, rich data structures
  (e.g., Protocol Buffers \cite{misc:protobuf}, an efficient binary-encoded,
  open-source data format widely used in the industry). To improve developer
  productivity, it is important to speed up the \emph{time-to-first-result},
  \emph{time-to-full-scale-result}, and \emph{time-to-trained-model} for
  machine learning. On the other hand, from a distributed systems standpoint,
  it is hard to simultaneously optimize for pipeline speed, resource cost, and
  reliability.

  \item How do we make a production cluster, hosting several large datasets
  with multiple developers simultaneously running pipelines, cost efficient by
  reducing the resource footprint? This is a common problem especially for
  developers in popular clusters (e.g., AWS, Azure, Google Cloud) who scale up
  (or down) their clusters for analytic workloads, because it is inefficient to
  dedicate a full cluster of machines and local storage. For example, consider
  a 20 TB dataset. We could use a dedicated cluster with 20 TB of RAM and local
  storage to fit the entire data in memory. However, it is about 5$\times$
  cheaper if we use a system with 2 TB of RAM, and about 40$\times$ cheaper if
  we use a system with 200 GB of RAM coupled with network storage
  \cite{misc:aws}. Moreover, the operational overhead with building and
  maintaining larger clusters is magnified as the memory requirements increase
  for petabyte scale datasets. As we see later, our system is built with these
  constraints in mind and offers good performance while being cost efficient.

\end{itemize}

In this paper, we discuss how WarpFlow addresses such challenges with below
features:

\begin{itemize}
  \item

  Supports fast, interactive data analytics to explore large, noisy
  spatiotemporal datasets with a pipeline-based query language using composite
  indices and spatiotemporal operators. The underlying techniques are easy to
  deploy in a cost-efficient manner on shared analytics clusters with datasets
  available on networked file systems such as in AWS, Microsoft Azure, Google
  Cloud \cite{misc:aws,misc:azure,misc:gcp}.

  \item

  Supports two complementary execution modes -- \emph{(i)} an ``insane''
  interactive mode, with an always-on speed optimized cluster and ``best
  effort'' machine failure tolerance, and \emph{(ii)} a batch mode for running
  large-scale queries with auto-scaling of resources and auto-recovery for
  reliable executions. In batch mode, WarpFlow automatically generates an
  equivalent Flume pipeline and executes it. In practice, we see development
  times are often 5 -- 10$\times$ faster than writing and building equivalent
  Flume pipelines from scratch, and helps our data developers gain a big
  productivity boost.

  \item

  Accelerates machine learning workflows by providing interfaces for faster
  data selection and built-in utilities to generate and process the training
  and test data, and enabling large-scale model application and inference.

\end{itemize}

The rest of the paper is structured as follows. In Section \ref{related_work},
we present related work and contrast our goals and assumptions. In Section
\ref{overview}, we give an overview of WarpFlow and its design choices. The
detailed architecture of WarpFlow and its components is presented in Section
\ref{architecture}. We present the application of WarpFlow to machine learning
use cases in Section \ref{machine_learning}. We present an example use case and
experimental results in Section \ref{experiments}, followed by conclusions in
Section \ref{conclusions}.

\section{Related Work} \label{related_work}

WarpFlow builds on years of prior work in relational and big data systems,
spatiotemporal data and indexing structures. In this section, we will summarize
the key common threads and differences. To the best of our knowledge, this is
the first system (and public description) that scales in practice for hundreds
of terabytes to petabytes of rapidly growing spatiotemporal datasets. While
there are large spatial image databases that store petabytes of raster images
of the earth and the universe (e.g., Google Earth), they of course have
different challenges (e.g., storing large hi-resolution images, and extracting
signals).

First, systems such as PostgreSQL \cite{misc:postgres}, MySQL \cite{misc:mysql}
offer a host of geospatial extensions (e.g., PostGIS \cite{misc:postgis}). To
tackle larger datasets on distributed clusters, recent analytical systems
propose novel extensions and specialized in-memory data structures (e.g., for
paths and trajectories) on Spark/Hadoop \cite{zaharia:spark} clusters
\cite{eldawy:spatialhadoop,xie:simba,xie:simbatraj,shang:dita,costa:spate,yu:geospark}.

Specifically, the techniques in \cite{xie:simba,xie:simbatraj,shang:dita} adopt
Spark's RDD model \cite{zaharia:spark} and extend it with a two-level indexing
structure. This helps prune RDD partitions but partitions containing matched
data need to be paged into memory for further filtering. These techniques work
well when (1) the data partition and indices fit in main memory on a
distributed cluster, (2) data overflows are paged into local disks on the
cluster, (3) the queries rely on the partition and block indices to retrieve
only relevant data partitions into available memory. In such cases, the
techniques work well to optimize CPU costs and can safely ignore IO costs in a
cluster. However, for our pipelines, we deal with numerous large datasets on a
shared cluster, so developers can run pipelines on these datasets concurrently.
We need to optimize both CPU and IO costs making use of fine-grained indexing
to selectively access the relevant data records, without first having to load
the partitions. As we see later, our techniques scale for multiple, large
datasets on networked file systems, while minimizing the resource footprint for
cost efficiency.

Second, WarpFlow supports two execution environments for pipelines. For long
running pipelines that need to deal with machine restarts and pipeline retries,
systems like MapReduce \cite{dean:mr}, Flume \cite{chambers:flume} and Spark
\cite{zaharia:spark} adopt checkpoint logs that allow a system to recover from
any state. For fast, interactive and short-running queries, systems like Dremel
\cite{melnik:dremel} drop this overhead, support an always running cluster and
push retries to the client applications. WarpFlow supports the best of both
worlds, by offering the developer two modes by relying on two separate
execution engines -- one for long running queries and one for fast, interactive
queries.

Third, how to express complex transformations on rich data has been an area of
active work, from SQL-like declarative formats to full procedural languages
(e.g., Scala, C++). For example, Shasta \cite{manoharan:shasta} uses RVL, a
SQL-like declarative language to simplify the queries. In contrast, WarpFlow's
language (WFL) uses a functional query language to express rich data
transformations from filters, aggregates, etc. to machine learning on tensors.
To speed up the iterations, WFL does not have a compilation step and interprets
the query dynamically at runtime. In addition, WFL is extensible to natively
support external C++ libraries such as TensorFlow \cite{misc:tensorflow}. This
makes it easy to integrate domain-specific functionalities for new types of
large-scale data analysis. Using a common WarpFlow runtime for data-intensive
operations speeds up the overall process, similar to Weld \cite{shoumik:weld}.
However, Weld integrates multiple libraries without changing their APIs by
using an optimized runtime, while WarpFlow provides a compatible API to use
libraries and their functions. Like F1 \cite{shute:f1}, WarpFlow uses Protocol
Buffers as first-class types, making it easy to support rich, hierarchical
data. Furthermore, WarpFlow uses Dynamic Protocol Buffers \cite{misc:dynproto},
so that developers can define and consume custom data structures \emph{on the
fly} within the query. As we see later, this helps developers to iterate on
pipelines without expensive build-compile cycles.

\section{Overview} \label{overview}

A WarpFlow pipeline consists of a data source that generates the \emph{flow} of
Protocol Buffers, followed by operators that transform the Protocol Buffers,
and the expressions to define these transformations. WarpFlow is fully
responsible for maintaining the state of the datasets including data sharding
and placement, distributing the pipeline, executing it across a large cluster
of machines, and load-balancing across multiple concurrent queries.

\begin{figure}[h]
	\centering
	\includegraphics[width=3.3in]{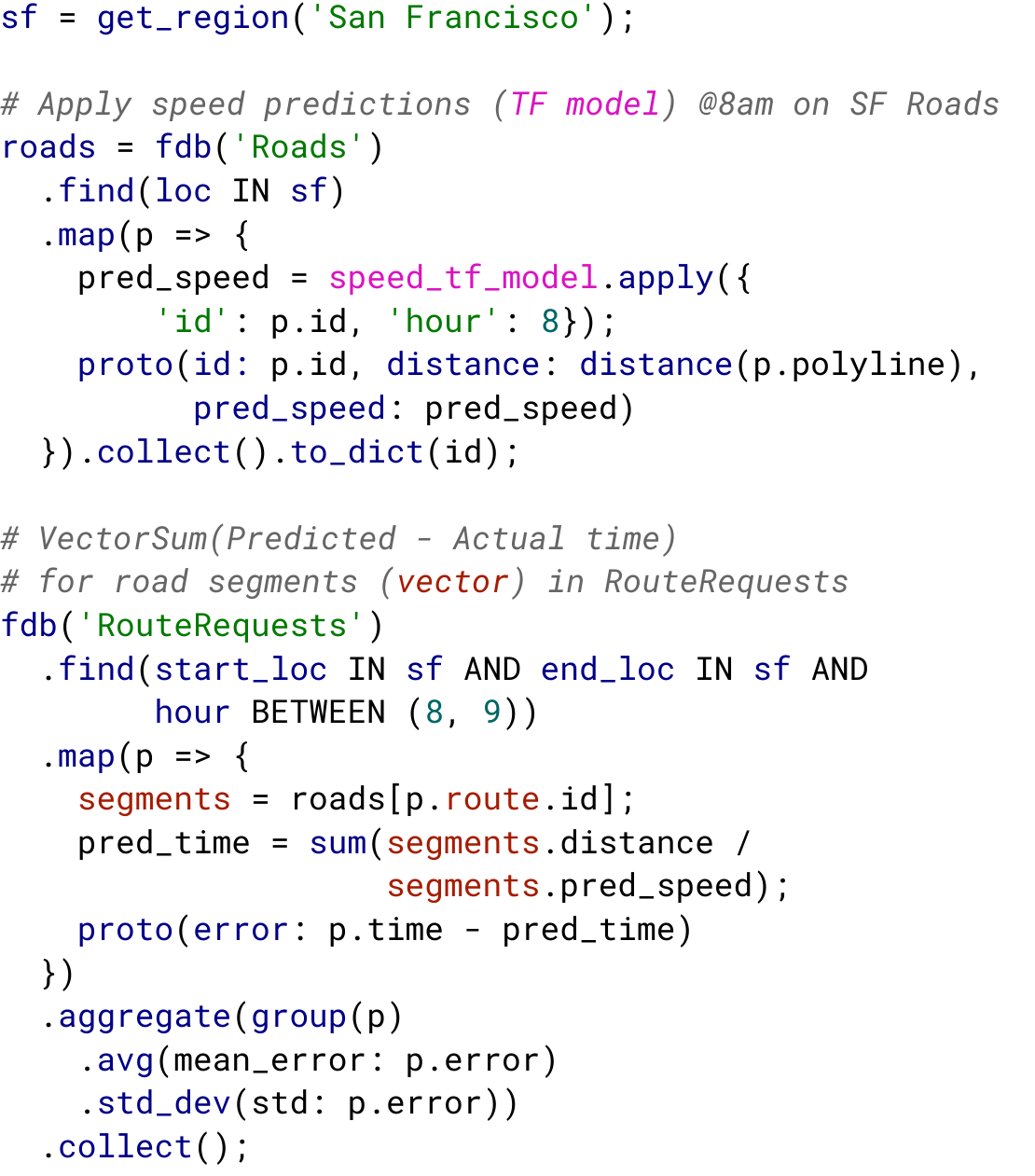}
	\caption{WFL snippet for the sample query.}
	\label{fig:sample_wfl_snippet}
\end{figure}

For example, consider how to evaluate the quality of a road speed prediction
model. To get a quick insight, we look at the routing requests in San Francisco
in the morning rush hour from 8\,am -- 9\,am. We start with the roads dataset --
a collection of roads along with their identifiers and geometry, and use its
indices to generate a stream of Protocol Buffers corresponding to roads in San
Francisco. For each road segment, we apply our speed prediction model to get
the predicted speed for 8\,am -- 9\,am, and compute the distance of the road
segment from its polyline. Next, we use the route requests dataset, which is a
collection of all routing requests along with the request time, the suggested
route and the actual travel time, and use its indices to get all relevant route
requests. We join the results with the road segment information from the
previous step, and use the distance and the predicted speed of the road
segments along the suggested route to compute the predicted travel time. The
difference in the predicted travel time and the actual travel time gives the
error in our prediction. Finally, we aggregate these measurements to get the
mean and standard deviation of the errors in travel time.

WarpFlow facilitates such common operations with a concise, functional
programming model. Figure \ref{fig:sample_wfl_snippet} shows a WFL snippet for
the previous example; the red fields are repeated (vectors) and the pink fields
are TensorFlow objects. Notice (1) index-based selections to selectively read
only the relevant records, and (2) operators for vectors, TensorFlow model
applications, and geospatial utilities. For example, the dictionary lookups
using a vector to get the road segments, the vector division to get travel
times per segment, and the geospatial utility for distance computation.

\subsection{Design choices}

\begin{figure}[ht]
	\centering
	\includegraphics[width=3.3in]{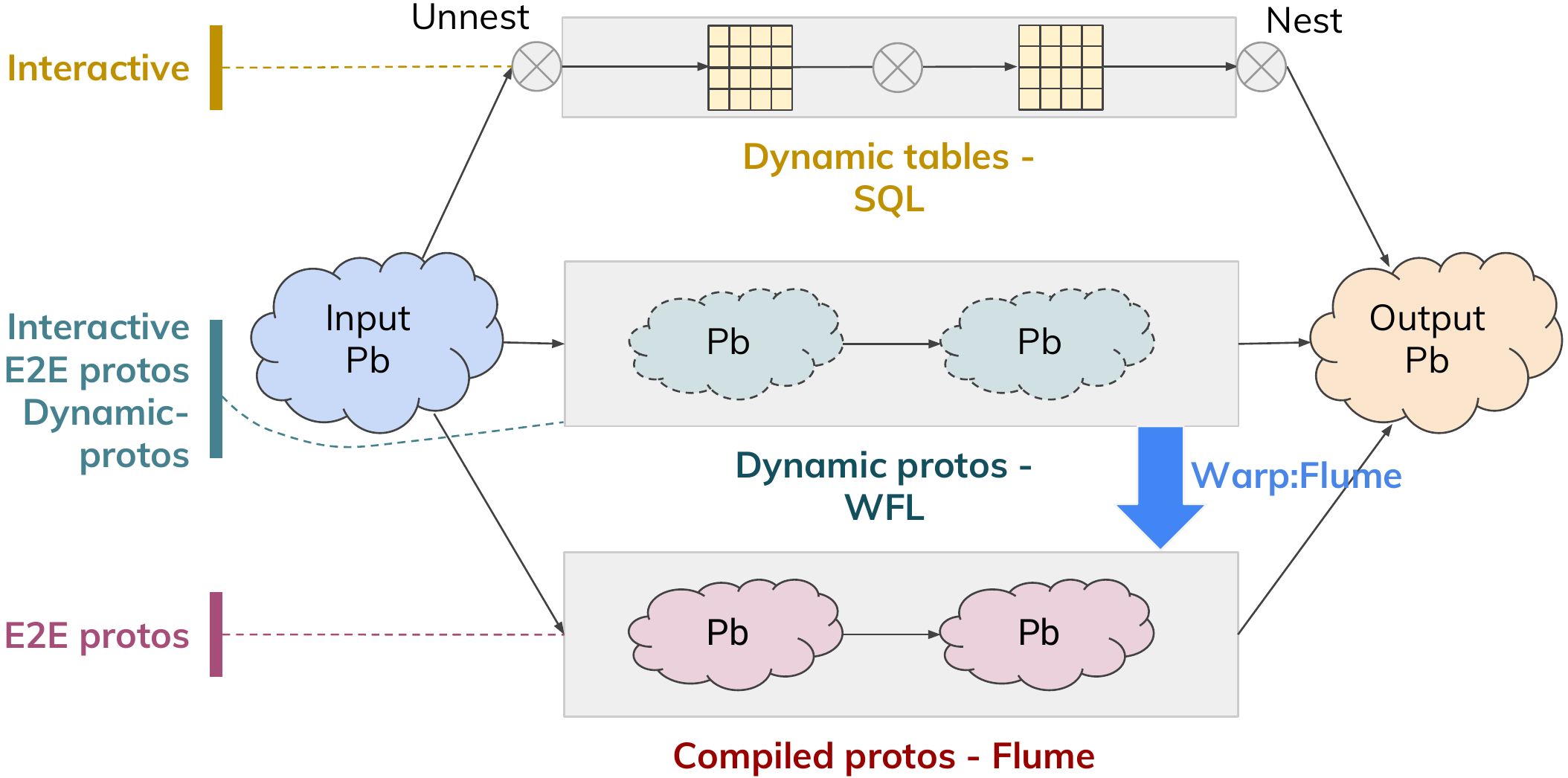}
	\caption{A simplified comparison of different data processing systems.}
	\label{fig:data_processing_options}
\end{figure}

Consider how to query and transform a Protocol Buffers dataset into a new
Protocol Buffers result set. Figure \ref{fig:data_processing_options} presents
a simplified conceptual depiction of different data processing systems, along
with some of the relevant benefits they offer (e.g., interactivity, end-to-end
Protocol Buffers, Dynamic Protocol Buffers, etc.).

\paragraph*{\textbf{Data pipeline model}} Data is usually modeled as relational
or hierarchical structures. Systems either: (a) retain Protocol Buffers and
manipulate them directly (e.g., MapReduce, Flume, Spark), or (b) re-model data
into relational tables (e.g., flatten repeated fields) and manipulate with
relational algebra (e.g., MySQL, Dremel, F1).

WarpFlow chooses (a): it retains Protocol Buffers at every stage of the
pipeline, for inputs, transforms, and outputs. Similar to Flume and Spark,
developers compose deep pipelines of maps, filters, and aggregates on these
hierarchical and nested structures.

\paragraph*{\textbf{Interactivity}} Interactivity is a highly desirable feature
for developers, often termed as REPL (read-evaluate-print-loop) in popular
frameworks like Python and Scala \cite{misc:repl}. Interactive data systems
enable developers to quickly compose/iterate and run queries, reducing the
time-to-first-result and speeding up the development and debug cycle with
instant feedback. Such systems are typically interpreted in nature as opposed
to being compiled, and have short runtimes to execute full and incremental
queries in a session.

WarpFlow offers a similar experience by making it easy for developers to (a)
access and operate on the data, and (b) iteratively build pipelines.
Specifically, it supports:

\begin{itemize}
  \item

  Always-on cluster for distributed execution of multiple ad hoc queries in
  parallel. Composite indices over hierarchical datasets and popular
  distributed join strategies \cite{hector:database} to help developers
  fine-tune queries, such as broadcast joins, hash joins and index-based joins.

  \item

  Query sessions to incrementally build and run queries with partial context
  kept in the cluster while the user refines the query. Also, full
  auto-complete support in query interfaces, not just for the language but also
  for the structure of the data, and the data values themselves.

  \item

  A strong toolkit of spatiotemporal functions to work with rich space-time
  datasets, and utilities to allow querying over a sample to quickly slice
  through huge datasets.

\end{itemize}

\section{Architecture} \label{architecture}

\begin{figure}[h]
  \centering
  \includegraphics[width=3.3in]{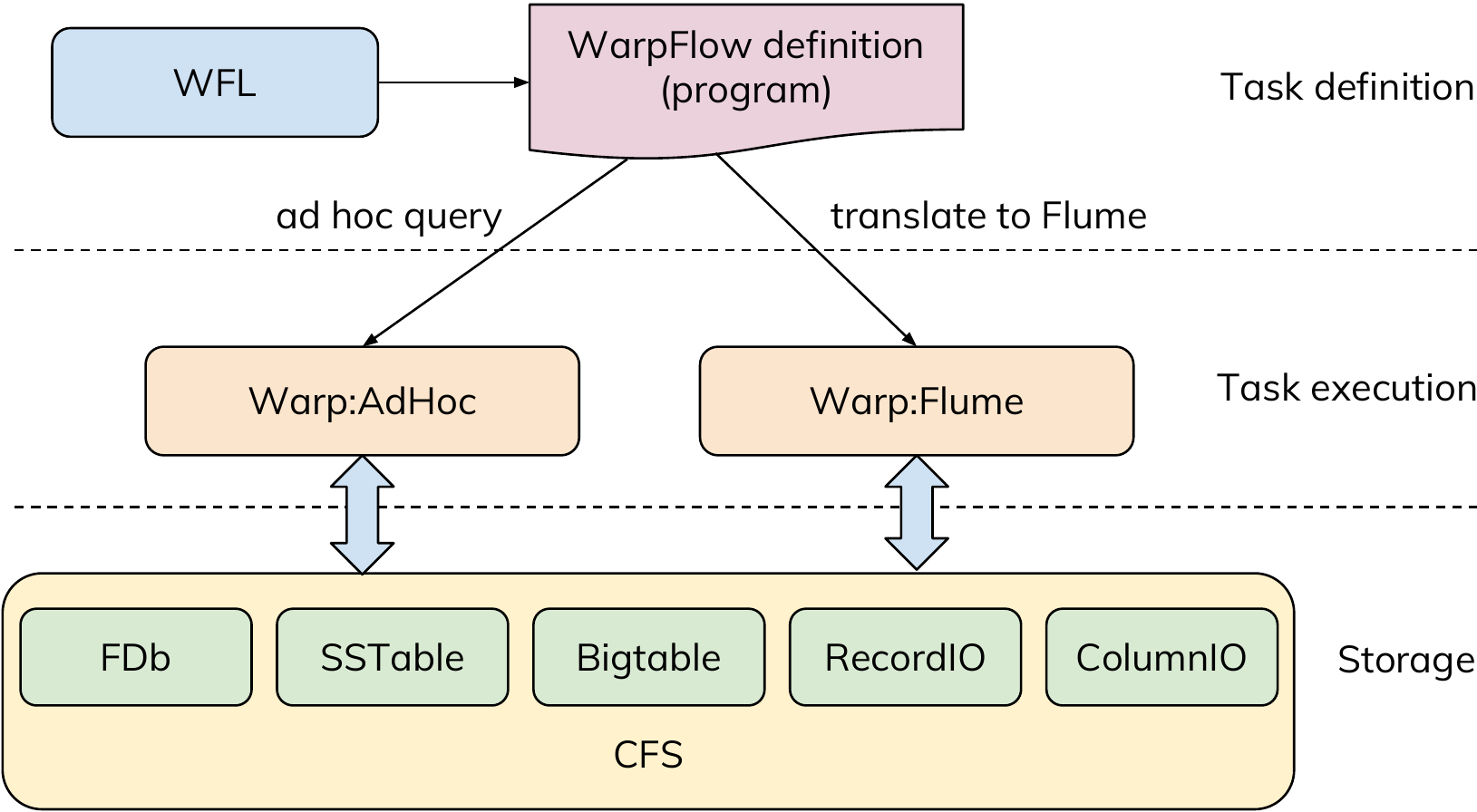}
  \caption{WarpFlow architecture.}
  \label{fig:flow_architecture}
\end{figure}

The WarpFlow system has three key components to handle the data storage, task
execution, and task definition functionalities, as shown in Figure
\ref{fig:flow_architecture}. The data storage layer holds the Protocol Buffers
data sources in one of several common storage formats. In addition, WarpFlow
builds custom ``FDb'' indices, optimized for indexing Protocol Buffers data.
The task execution layer reads the Protocol Buffers data from the storage layer
and carries out the query execution. WarpFlow supports two execution modes. The
pipelines run in an interactive, ad hoc manner using ``Warp:AdHoc'', or as a
batch job on Flume using ``Warp:Flume''. The specification of the query and its
translation to the execution workflow is carried out by the task definition
layer. WarpFlow uses a functional query language, called ``WarpFlow Language''
(WFL), to express the queries concisely.

In this section, we describe each of the key components in detail.

\subsection{Data storage and indexing} \label{data_storage}

In the following sections, we describe the underlying structure and layout of
FDb, and the different index types that are available.

\begin{figure*}[ht]
	\centering
	\includegraphics[width=5in]{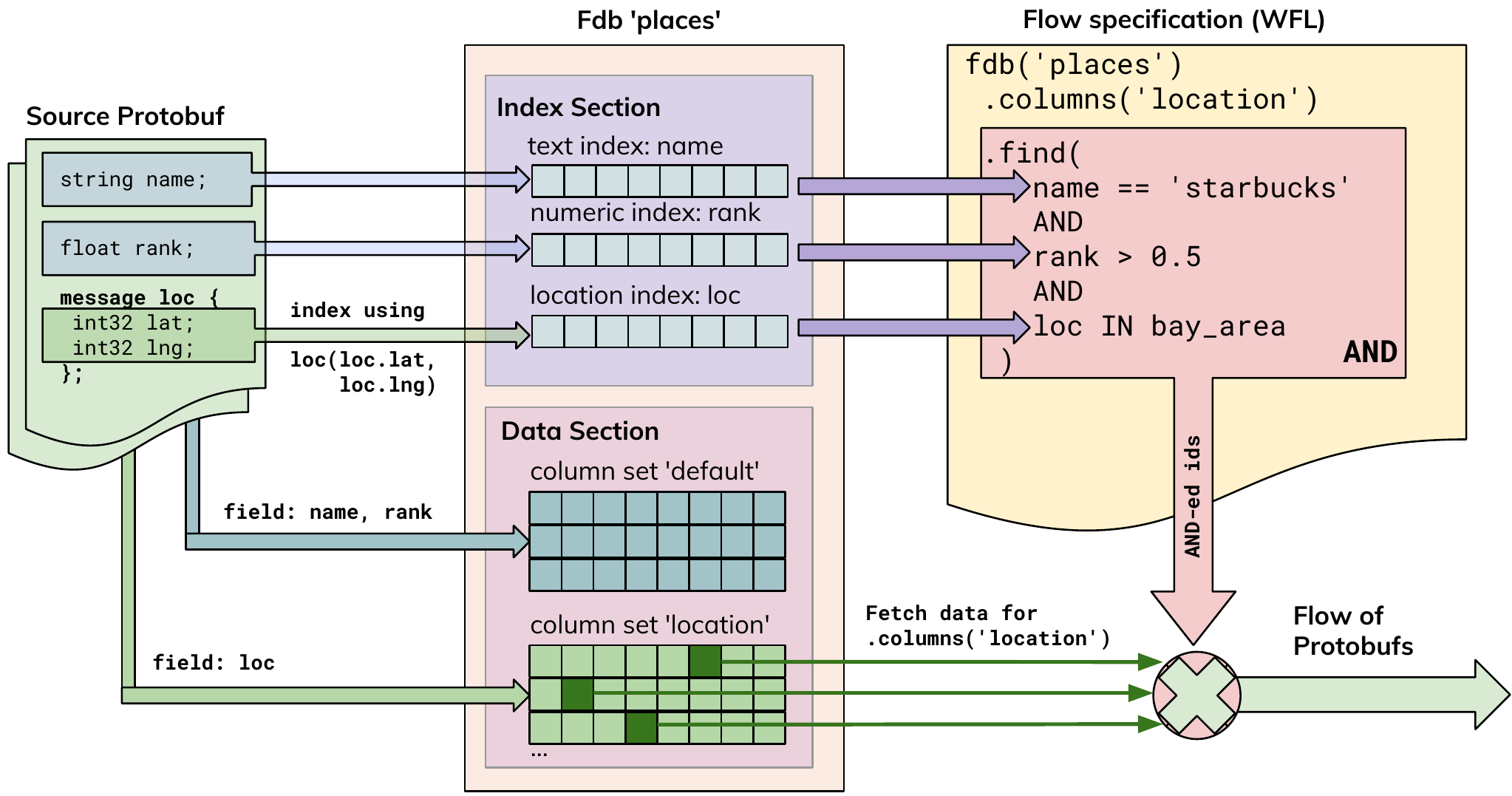}
	\caption{Data layout of a sample FDb showing various indices and column set
		data storage, along with an example query showing index-based data
		selection.}
	\label{fig:sample_fdb}
\end{figure*}

\subsubsection{FDb: Indexing and storage format} \label{fdb_format}

FDb is a column-first storage and search database for Protocol Buffers data.
Like many distributed data formats, FDb is a sharded storage format, sharding
both the data and the indices. Each FDb shard stores data values organized by
\emph{column sets} (similar to column families in Bigtable
\cite{chang:bigtable}) and index values mapped to document-ids within the
shard.

The basic FDb layout is illustrated in Figure \ref{fig:sample_fdb} showing a
sample Protocol Buffers record with fields \texttt{name}, \texttt{rank} and a
sub-message \texttt{loc} with fields \texttt{lat} and \texttt{lng}. The data
layout has separate sections for indices and data. The data section is further
partitioned into column sets. A sample query, shown on the right in Figure
\ref{fig:sample_fdb}, accesses the necessary indices to narrow down the set of
documents to read, which are then read column-wise from the column sets.

FDb is built on a simple key-value storage abstraction and can be implemented
on any storage infrastructure that supports key-value based storage and lookups
e.g., LevelDb \cite{misc:leveldb}, SSTable \cite{chang:bigtable}, Bigtable, and
in-memory datastores. We use SSTables to store and serve our ingested datasets
as read-only FDbs. The sorted key property is used to guarantee iteration-order
during full table scans and lookups. We implement read-write FDbs on Bigtable
for streaming FDbs, including for query profiling and data ingestion logs.

\subsubsection{Index types} \label{index_types}

FDb supports indexing a variety of complex value types. A single field in the
Protocol Buffers (e.g., navigation path) can have multiple indices of different
types. In addition to basic inverted text and range indices for text and
numeric values, FDb supports geometry based indices e.g., locations, areas,
segments, etc. as described below.

\paragraph*{\textbf{location}} Location indices are intended for data fields
that represent punctual places on the Earth, typically identified by latitude
and longitude. Internally, we use an integer-representation of the location's
Mercator projection \cite{maling:coord} with a precision of several
centimeters. As such, locations to the north of $85\degree$ N and south of
$85\degree$ S are not indexable without some translation. The selection of
location index can be specified by a bounding box (with south-west and
north-east corners) or by a region (e.g., a polygonal area) to fetch all the
documents where the location field is within the given region.

\begin{figure}[h]
	\centering
	\includegraphics[width=3in]{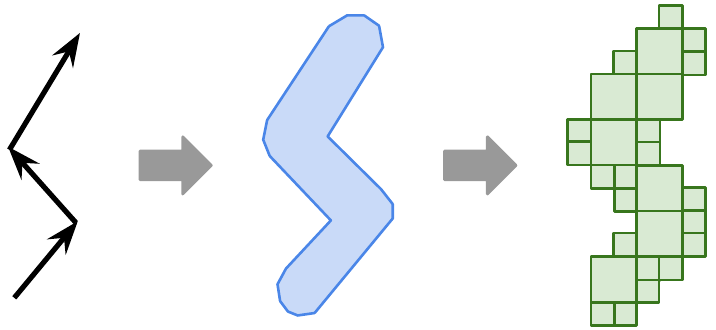}
	\caption{Indexing a path using an area tree.}
	\label{fig:path_index}
\end{figure}

\paragraph*{\textbf{area}} Area index is used for geospatial regions, usually
represented by one or more polygons. We use \emph{area trees} to represent
these indices. The selection of areas can be made efficiently, either by a set
of points (i.e., all areas that cover these points) or by a region (i.e., all
areas that intersect this region). In addition, these indices are also used to
index geometries other than regions by converting them to \emph{representative
areas}. For example, a point is converted to an area by expanding it into a
circular region of a given radius; a path is converted to an area by expanding
it into a strip of a given width. These areas are then indexed using an area
tree, as shown in Figure \ref{fig:path_index}. This enables indexing richer
geospatial features such as navigation paths, parking and event zones.

Area trees are very similar to quad-trees \cite{finkel:quadtree}; the main
difference is that the space of each node is split into 64 ($8\times8$)
sub-nodes as opposed to four ($2\times2$) sub-nodes in a quad-tree. The 64-way
split of each node leads to an efficient implementation and maps naturally to
the gridding of the Earth in the spherical Mercator projection. They can be
combined (union, difference or intersection) in a fast, efficient manner, and
can be made as fine as needed for the desired granularity (a single pixel can
represent up to a couple of meters). In addition to indexing purposes, they are
used for representing and processing geospatial areas in a query.

Indices and column sets are annotated on the Protocol Buffers specification
using field options. For any field or sub-field in the message, we use options
to annotate it with the index type (e.g., options
\texttt{\detokenize{index_text}} and \texttt{\detokenize{index_tag}} to create
text and tag indices). We also define the mapping of fields to column sets. In
addition, we can define virtual fields for the purpose of indexing. These
\emph{extra fields} are not materialized and are only used to create the index
structure.

\subsubsection{Data de-noising} \label{data_denoising}

\begin{figure}[h]
	\centering
	\includegraphics[width=3.3in]{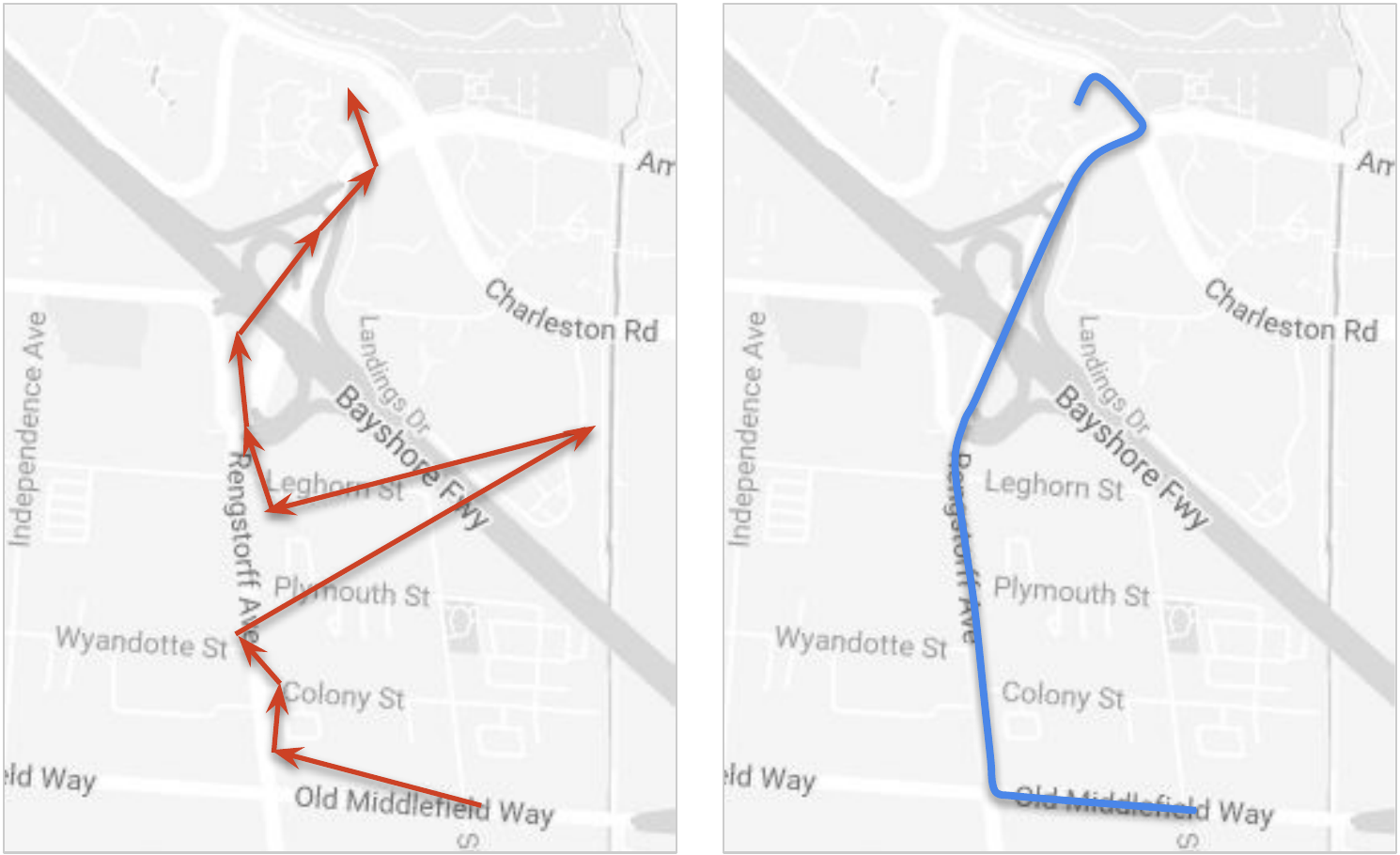}
	\caption{A noisy trace (left) that is snapped to a road route (right).}
	\label{fig:noisy_path}
\end{figure}

As mentioned earlier, spatiotemporal data from devices often have poor location
accuracy or occasional bad network links. Our pipelines need to be resilient to
noisy data, and should filter and smooth the data. The presence of noise
transforms a precise location or a path into a probabilistic structure
indicating the likely location or path. WarpFlow provides methods to construct
and work with probabilistic representations of the spatial data, and to project
and snap them to a well-defined space. For example, we can snap a noisy
location to a known point-of-interest (POI), or snap a noisy navigation path
with jittered waypoints to a smooth route along road segments, as shown in
Figure \ref{fig:noisy_path}. Snapping is often done by selecting the fuzzy
regions of interest and applying a machine-learned (ML) model using signals
such as the popularity of places and roads, similarity to other crowdsourced
data, and suggested routes from a routing engine (as we see later, WarpFlow
supports TensorFlow to train and apply ML models).

Area indices help us work with such noisy geospatial data and snappings.
Representative areas are a natural way to identify probabilistic locations and
paths. For example, a probabilistic location can be represented by a mean
location and a confidence radius (i.e., a circular region) depending on the
strength of the noise. Similarly, a probabilistic path can be represented by a
curvilinear strip whose thickness indicates the strength of the noise. Recall
that this area is not a bounding box of the points in a path, but an envelope
around the path so time ordering is preserved. We can then use this fuzzy
selection of data and intersect with filter conditions. As we see later, these
simple, fuzzy selections help us handle large datasets in a cost-efficient
fashion on shared cloud clusters.

\subsection{WarpFlow language (WFL)}

WarpFlow uses a custom, functional programming language, called WarpFlow
language (WFL), to define query pipelines with hierarchical datasets. A common
problem when working with deeply nested, hierarchical data structures (e.g.,
Protocol Buffers) is how to (1) first express the query pipelines, and (2)
later, evolve the pipelines as underlying structures and requirements change.
To address this problem, WFL draws inspiration from modern, functional
languages, such as Scala \cite{misc:scala} that draw on decades of software
engineering best practices for code structuring, maintenance, and evolution.
WFL adopts two key elements for succinct queries: (1) a pipeline-based approach
to transform data with sequentially chained operations, and (2) hierarchical
structures as primitive data types, so operators work on vectors, tensors and
geospatial structures.

The full language definition and its constructs are out of the scope of this
paper. Instead, we present a simplified overview of the language in this
section.

\subsubsection{Data types}

Protocol Buffers objects in WFL have two properties: (1) \emph{type:} the data
type of the object, which is one of \{bool, int, uint, float, double, string,
message\}. (2) \emph{cardinality:} the multiplicity of the object i.e.,
singular or repeated; singular and repeated fields are treated as scalars and
vectors respectively for most operations.

In addition to the basic data types, WFL provides utilities such as
\texttt{array}, \texttt{set}, \texttt{dict} and geo-utilities (e.g.,
\texttt{point}, \texttt{area}, \texttt{polygon}, etc.) to compose basic data
types into pre-defined, higher-level objects.

\subsubsection{Operators and expressions}

Each stage in a WFL pipeline generates a series of Protocol Buffers records
with the same structure. We call this series a \emph{flow} of Protocol Buffers.
Flow provides operators that transform the underlying Protocol Buffers. Most of
these operators accept a \emph{function expression} as an argument, that
defines the transformation to be performed. Each operator, in turn, generates a
new flow of transformed Protocol Buffers. A typical WFL pipeline with several
chained operators has the following syntax:

\begin{verbatim}
flow
  .<flow_operator_1>(p => {body_1})
  .<flow_operator_2>(p => {body_2})
  .<flow_operator_3>(p => {body_3})
\end{verbatim}

A transformation is defined using expressions composed of the aforementioned
data types, operators, and higher-order functions, like in any programming
language. The expression body does not have a return statement; the final
statement in the body becomes its return value. Each flow operator may require
a specific return type, e.g., a \texttt{filter} operator expects a boolean
return type and a \texttt{map} operator expects a Protocol Buffers return type.
The primary flow operators and their functions are presented in Table
\ref{tab:flow_operators}.

\begin{table}[ht]
  \begin{center}
  \begin{tabular}{ p{0.6in} p{2.4in} }
    \toprule
    \textbf{Operator} & \textbf{Function} \\

    \midrule
    \multicolumn{2}{l}{\emph{Basic transformations}} \\
    \addlinespace
    \texttt{map} &
      Transform a Protocol Buffers record into another Protocol Buffers
      record. \\
    \texttt{filter} &
      Filter records in the flow based on boolean condition. \\
    \texttt{flatten} &
      Flatten repeated fields within Protocol Buffers into multiple Protocol
      Buffers. \\

    \midrule
    \multicolumn{2}{l}{\emph{Reordering a flow}} \\
    \addlinespace
    \texttt{\detokenize{sort_asc}}, \texttt{\detokenize{sort_desc}} &
      Sort the flow (in ascending or descending order) using an expression. \\

    \midrule
    \multicolumn{2}{l}{\emph{Resizing a flow}} \\
    \addlinespace
    \texttt{limit} &
      Limit the number of records in the flow. \\
    \texttt{distinct} &
      Remove duplicate records from the flow, based on an expression. \\
    \texttt{aggregate} &
      Aggregate the records in the flow, possibly grouping them by one or more
      fields or expressions, using predefined aggregates (e.g., count, sum,
      avg). \\

    \midrule
    \multicolumn{2}{l}{\emph{Merging flows}} \\
    \addlinespace
    \texttt{join} &
      Merge Protocol Buffers from two different flows using a hash join. \\
    \texttt{\detokenize{sub_flow}} &
      Join Protocol Buffers from the flow with a sub-flow using index join. \\

    \midrule
    \multicolumn{2}{l}{\emph{Materializing a flow}} \\
    \addlinespace
    \texttt{collect} &
      Collect the Protocol Buffers records from the flow into a variable. \\
    \texttt{save}, \texttt{\detokenize{to_sstable}},
      \texttt{\detokenize{to_recordio}} &
      Saves the Protocol Buffers from the flow to FDb, SSTable or RecordIO
      \cite{misc:recordio}. \\
    \bottomrule
  \end{tabular}
  \caption{Primary flow operators.}
  \label{tab:flow_operators}
  \end{center}
\end{table}

Expressions compose data types using simple operators, e.g., \texttt{+},
\texttt{-}, \texttt{*}, \texttt{/}, \texttt{\%}, \texttt{AND}, \texttt{OR},
\texttt{IN}, \texttt{BETWEEN}. These operators are overloaded to perform the
corresponding operation depending on the type of the operands. In addition to
these simple operators, a collection of utilities make it easier to define the
transformations.

Furthermore, WFL seamlessly extends the support for these operations to
repeated data types. If one or more of the operands is of a repeated data type,
the operation is extended to every single element within the operand. This
greatly simplifies the expressions when working with vectors, without having to
iterate over them.

Finally, WFL offers a large collection of utilities to simplify and help with
common data analysis tasks. Besides basic toolkits to work with strings, dates
and timestamps, it provides advanced structures such as HyperLogLog sketches
for cardinality estimation of big data \cite{flajolet:hll}, Bloom filters
\cite{bloom:filters} for membership tests, and interval trees
\cite{cormen:algo} for windowing queries. It also has a geospatial toolkit for
common operations such as geocoding, reverse-geocoding, address resolution,
distance estimation, projections, routing, etc.

\subsubsection{Extensibility}

In addition to the built-in features and functions, WFL is designed to be an
extensible language. It allows custom function definitions within WFL. It also
features a modular function namespace for loading predefined WFL modules.
Furthermore, we can easily extend the language by wrapping the APIs for
third-party C++ libraries and exposing them through WFL. We use this approach
to extend WFL to support TensorFlow \cite{misc:tensorflow} API, enabling
machine learning workflows with big data (see Section \ref{machine_learning}
for more details). This capability elevates the scope of WarpFlow by providing
access to the vast body of third-party work.

\subsection{WarpFlow execution}

WarpFlow pipelines can be executed as interactive, ad hoc jobs with Warp:AdHoc,
or as offline, batch jobs with Warp:Flume depending on the execution
requirements. Developers typically use Warp:AdHoc for initial data explorations
and quick feedback on sampled datasets. The same WFL pipeline can later be
executed on full-scale datasets as a batch job on Flume using Warp:Flume. In
this section, we first describe Warp:AdHoc and its underlying components. The
logical model of data processing is maintained when converting WFL pipelines to
Flume jobs using Warp:Flume.

\subsubsection{Warp:AdHoc}

Warp:AdHoc is an interactive execution system for WFL pipelines. A query
specification from WFL is translated into a directed acyclic graph (DAG)
representing the sequence of operations and the flow of Protocol Buffers
objects from one stage to the next. Warp:AdHoc performs some basic
optimizations and rewrites to produce an optimized DAG. The execution system is
responsible for running the job as specified by this DAG.

\begin{figure}[h]
  \centering
  \includegraphics[width=3.3in]{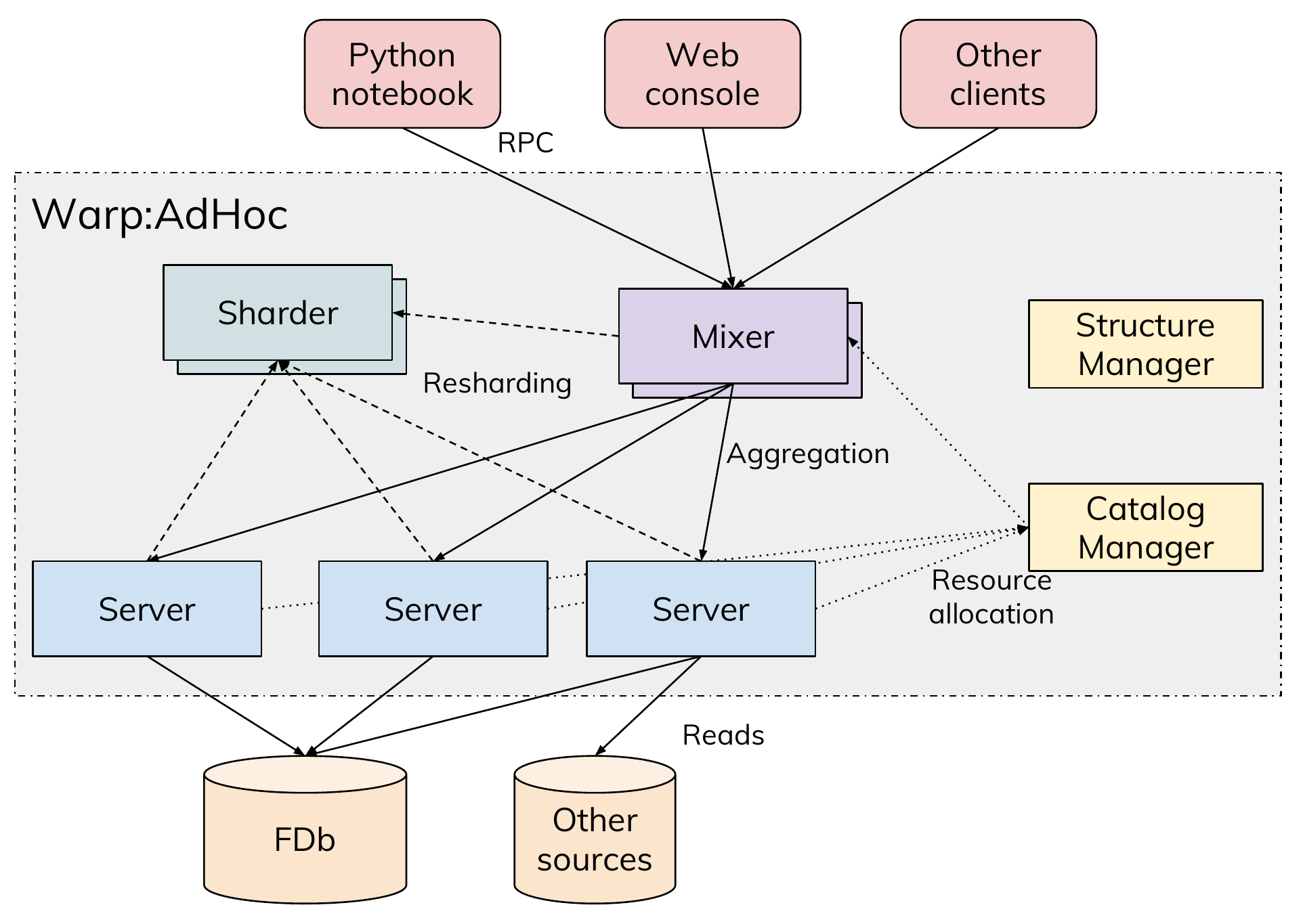}
  \caption{Warp:AdHoc architecture.}
  \label{fig:warp_exec_architecture}
\end{figure}

The high-level architecture of Warp:AdHoc is shown in Figure
\ref{fig:warp_exec_architecture}. Developers work on clients such as an
interactive Python notebook (e.g., Jupyter \cite{misc:jupyter}, Colaboratory
\cite{misc:colab}), or a web-based console, which interact with Warp:AdHoc via
a remote interface. Through this interface, clients communicate with a
\emph{Mixer}, which coordinates query execution, accumulates and returns the
results. The Mixer distributes the execution to \emph{Servers} and
\emph{Sharders} depending on the query.

The system state of Warp:AdHoc is maintained by a few metadata managers.
Specifically, \emph{Structure manager} maintains a global repository of
Protocol Buffers structures defined statically or registered at run-time.
\emph{Catalog manager} maintains pointers to all registered FDbs, and maps them
to Servers for query and load distribution.

\subsubsection{Dataset structures}

Although Warp:AdHoc can read and transform data from a variety of sources,
queries executed over them are typically slower due to full-scan of data. FDb
is the preferred input data source for Warp:AdHoc, where indices over one or
more columns can be used in \texttt{find()} to only read relevant data.

Warp:AdHoc needs to know the structure of the Protocol Buffers representing the
underlying data. For non-FDb data sources, these are provided by the developer.
For FDb sources these structures are typically registered using the name of the
Protocol Buffers with the Structure manager, and are referred to directly by
these names in WFL pipelines. These structures can be added, updated, or
deleted from the Structure manager as necessary.

\subsubsection{Dynamic Protocol Buffers}

SQL-based systems like Dremel and F1 enable fast, interactive REPL analysis by
supporting dynamic tables. Each SQL SELECT clause creates a new table type, and
multiple SELECTs can be combined into arbitrarily complex queries. However,
users do not need to define schemas for all these tables -- they are created
dynamically by the system.

Similarly, WarpFlow uses Dynamic Protocol Buffers \cite{misc:dynproto} to
provide REPL analysis. WFL pipelines define multi-step data transformations,
and the Protocol Buffers schema for each stage is created dynamically by the
system using Dynamic Protocol Buffers.

\begin{figure}[h]
  \centering
  \includegraphics[width=3.3in]{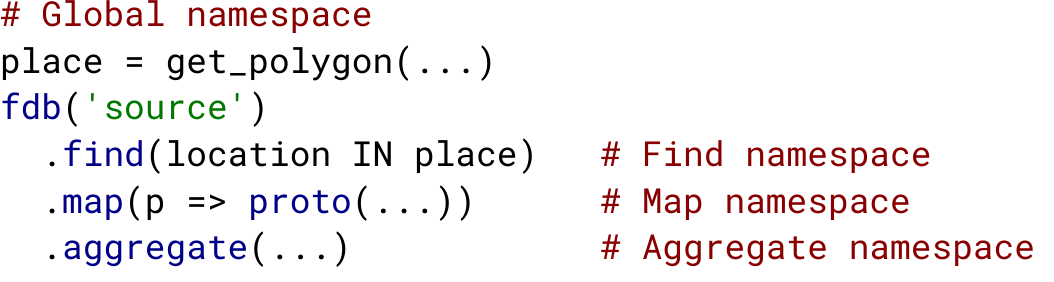}
  \caption{A sample WFL query with different Protocol Buffers structure after
    each stage.}
  \label{fig:dynamic_protobuf_snippet}
\end{figure}

For instance, in the sample WFL pipeline in Figure
\ref{fig:dynamic_protobuf_snippet}, the output of each stage (e.g.,
\texttt{map}, \texttt{aggregate}) has a Protocol Buffers structure that is
different from that of the source, and the necessary Protocol Buffers schemas
are implicitly defined by the system.

In addition, a WFL pipeline has a global namespace and a tree of local
namespaces corresponding to pipeline stages and functions. Each namespace has
variables with static types inferred from assignments. Values in these
namespaces can be used in Protocol Buffers transformations, as shown in the
\texttt{find()} clause in Figure \ref{fig:dynamic_protobuf_snippet}. These
namespaces are also represented by Dynamic Protocol Buffers with fields
corresponding to their defined variables.

In a relational data model, the data is usually normalized across multiple
tables with foreign-key relationships, reducing the schema size of an
individual table. In hierarchical datasets, the entire nested data is typically
stored together. Sometimes, the structure of this data has a deep hierarchy
that recursively loads additional structures, resulting in a schema tree with a
few million nodes. Loading the entire schema tree to read a few fields in the
source data is not only redundant but also has performance implications for
interactive features (e.g., autocompletion).

Instead, WarpFlow generates the \emph{minimal viable schema} by pruning the
original Protocol Buffers structure tree to the smallest set of nodes needed
for the query at hand (e.g., tens of nodes as opposed to millions of nodes). It
then composes a new Dynamic Protocol Buffers structure with the minimal viable
schema which is used to read the source data. This reduces the complexity of
reading the source data and improves the performance of interactive features.

\subsubsection{Query planning}

When a WFL query is submitted to Warp:AdHoc, a query plan is formulated to
determine pipeline distribution and resource requirements, similar to
distributed database query optimizers \cite{hector:database}. Most stages in
the pipeline are executed remotely on the Servers, followed by an optional
final aggregation on the Mixer. Query planning involves determining stages of
the pipeline that are remotely executed, the actual shards of the original data
source that are required for the query, and the assignment of execution
machines to these shards. Depending on the query, the planning phase also
optimizes certain aspects of the execution. For example, a query involving an
aggregation by a data sharding key is fully executed remotely without the need
for a final aggregation on the Mixer.

Query planning also determines the Protocol Buffers structures at different
stages in the pipeline. The structures for parts of the query that are executed
remotely are distributed to the respective Servers and Sharders.

\begin{figure}[h]
  \centering
  \includegraphics[width=3.3in]{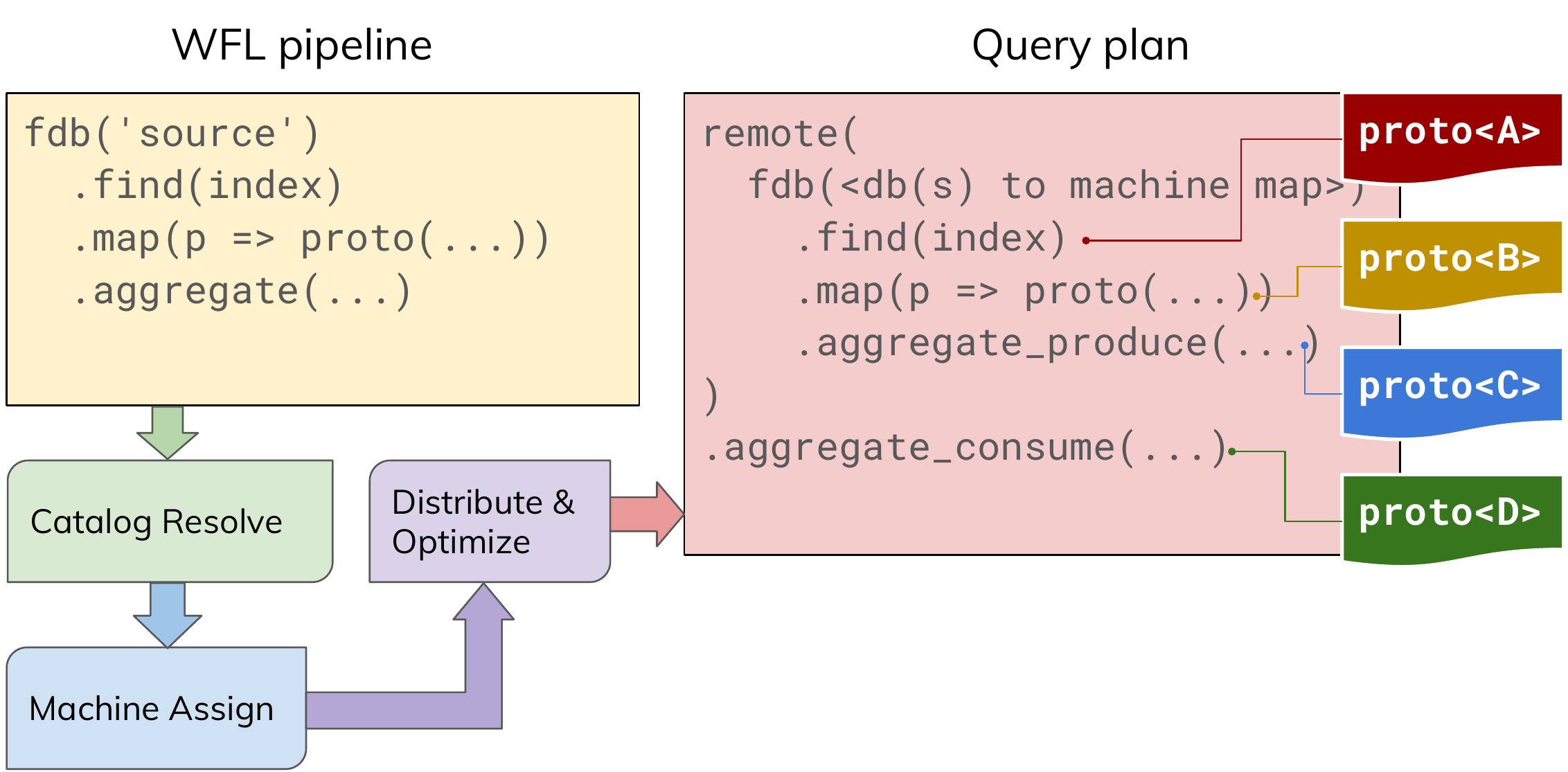}
  \caption{A Warp:AdHoc query plan showing the intermediate Protocol Buffers
    structures.}
  \label{fig:warp_exec_query_plan}
\end{figure}

The query plan for a typical WFL query is shown in Figure
\ref{fig:warp_exec_query_plan}. The pipelines within \texttt{remote(...)}
execute on individual Servers reading data from the assigned FDb shards from a
common file system. The \texttt{\detokenize{aggregate_consume(...)}} stage
aggregates partial results received from all the remote pipelines on the Mixer.

Data is transformed through different Protocol Buffers structures as it passes
through different stages in the pipelines. For example, in Figure
\ref{fig:warp_exec_query_plan}, \texttt{proto<A>} is the structure of the FDb
source data while \texttt{proto<B>}, \texttt{proto<C>} and \texttt{proto<D>}
are the structures of the output of \texttt{map()},
\texttt{\detokenize{aggregate_produce()}} and
\texttt{\detokenize{aggregate_consume()}} stages respectively.

\subsubsection{Distributed execution}

Warp:AdHoc executes WFL pipelines using a distributed cluster of Servers and
Sharders. Each WFL query requires the necessary resources (execution servers)
to be allocated before its execution begins. These resources are requested from
the Catalog manager, after the query planning phase. If resources are not
immediately available then the query waits in a queue until they are allocated.
Eventually, the Catalog manager allocates the Servers for execution, along with
assigning a subset of FDb shards to each Server for local reads and
transformations.

The query execution starts by setting up the corresponding pipeline stages on
Servers, Sharders and the Mixer. Servers then start reading from their assigned
FDb shards, transform the data as necessary, and return partial results to the
Mixer. Sharders perform intermediate shuffles and joins as specified by the
pipeline. The Mixer pipeline aggregates the partial results and returns the
final results to the client. To reclaim the resources when a query is blocked,
a time limit is imposed at various stages and its execution is re-attempted or
aborted.

WarpFlow makes it easy to query enormous amounts of data, for concurrent
queries. It offers \emph{execution isolation} -- each query gets its own
dedicated micro-cluster of Servers and Sharders for the duration of its
execution. This ensures that queries do not interfere with each other.

\subsubsection{Warp:Flume}

In addition to interactive execution, WarpFlow can automatically translate WFL
queries to Flume jobs for large-scale, batch processing. Warp:Flume is the
component of WarpFlow that is responsible for this translation and execution.

Each stage of a WFL pipeline is internally implemented using \emph{processors},
such as \texttt{find processor} for \texttt{find()}, \texttt{map processor} for
\texttt{map()}, and so on. To enable the translation to Flume jobs, each
processor is wrapped into a Flume function. In addition to these processors, we
also implement specialized Flume data readers that can work with FDb data
sources and use index selection for data fetching. The data processed by a
pipeline stage is wrapped into standard Flume data types such as
\texttt{flume::PCollection} and \texttt{flume::PTable<K,V>}
\cite{chambers:flume}, depending on the type of the processor.

Warp:AdHoc uses Dynamic Protocol Buffers to pass data between the stages. For
Warp:Flume, we use two ways to pass the data between the stages: \emph{(i)}
String encoding -- convert all the Protocol Buffers to strings, pass the string
data to the next stage, which then deserializes them into Protocol Buffers;
\emph{(ii)} Protocol Buffers encoding -- we retain the data as Protocol Buffers
and share the pointers between the stages, along with custom coders to process
these Dynamic Protocol Buffers. From our experiments, we notice that option
\emph{(i)} tends to be faster for simple pipelines with few stages where the
encoding and decoding overhead is minimal, but option \emph{(ii)} is faster for
longer pipelines with heavy processing and multiple stages. In general, we
notice a 25\% performance penalty when compared with an equivalent,
hand-written Flume job. Nevertheless, with Warp:Flume we typically see
development times are faster by about 5 -- 10$\times$. We believe the speed up
in the development time more than compensates for the small overhead in
runtimes.

\section{Machine learning} \label{machine_learning}

Machine learning (ML) brings novel ways to solve complex, hard-to-model
problems. With WarpFlow's extensible data pipeline model, we support TensorFlow
\cite{misc:tensorflow} as another pipeline operator extension for common use
cases. A typical workflow of an ML developer has the following main steps:

\begin{enumerate}
  \item

  Design a prototype ML model with an input feature set to provide an output
  (e.g., estimations or classifications) towards solving a problem.

  \item

  Collect labeled training, validation, and test data.

  \item

  Train the model, validate and evaluate it.

  \item

  Use the saved model to run large-scale inference or a large-scale evaluation.

\end{enumerate}

Usually, steps 1 -- 3 are repeated in the process of iterative model refinement
and development. A lot of developer time is spent in feature engineering and
refining the model so it is best able to produce the desired outputs. Each
iteration involves fetching training, validation and test data that make up
these features. In some cases, we see wait times of a few hours just to extract
such data from large datasets.

Quick turn around times in these steps accelerate the development and enable
testing richer feature combinations. Towards this end, WFL is extended to
natively support TensorFlow APIs for operations related to basic tensor
processing, model loading and model inferences.

To be able to fetch the data and extract the features from it, a set of basic
tensor processing operations are provided through TensorFlow utilities in WFL.
This minimal set of operations enables basic tensor processing and marshaling
that is needed to compose and generate features from a WFL query.

After getting hold of the data and the features, the training is performed
independently, usually in Python. While training within WarpFlow is possible,
it is often convenient to use frameworks that are specifically designed for
accelerated distributed training on specialized hardware \cite{misc:tpu}.
WarpFlow helps the developers get to the training faster through easy data
extraction.

With the completion of the training phase, there are the following use cases
for model application:

\begin{itemize}
  \item

  Trained models are typically evaluated on small test datasets, after the
  training cycle. However, the model's performance often needs to be evaluated
  on much bigger subsets of the data to understand the limitations that might
  be helpful in a future iteration.

  \item

  Once a model's performance has been reasonably tested, the model can be used
  for its inference -- predictions or estimations of the desired quantities. A
  common use case is to run the model offline on a large dataset and annotate
  it with the inferences produced by the model. For example, a model trained to
  identify traffic patterns on roads can be applied offline on all the roads
  and annotate their profile with predicted traffic patterns; this can later be
  used for real-time traffic predictions and rerouting.

  \item

  As an alternative to offline application, the inferences of the model can be
  used in a subsequent query that is computing derived estimates i.e., the
  model is applied online and its inferences are fed to the query.

\end{itemize}

To enable the above use cases, a set of TensorFlow utilities related to model
loading and application are added to WFL. For easier interoperability with
other systems, these utilities are made compatible with the standard SavedModel
API \cite{misc:savedmodel}.

\section{Experiments} \label{experiments}

A billion people around the world use Google Maps for its navigation and
traffic speed services, finding their way along 1 billion kilometers each day.
As part of this, engineers are constantly refining signals for routing and
traffic speed models. Consider one such ad hoc query: ``Which roads have highly
variable traffic speeds during weekday mornings?'' In this section, we evaluate
the performance of WarpFlow under different criteria for such queries.

For this paper, we use the following dataset and experimental setup to
highlight a few tradeoffs.

\begin{itemize}

  \item Google Maps maintains a large dataset of road segments along with their
  features and geometry, for most parts of the world. It also records traffic
  speed observations on these road segments and maintains a time series of the
  speeds. For these experiments, we use only the relevant subset of columns
  (observations and speeds) for past 6 months ($\sim$ 27 TB). For this dataset,
  we want to accumulate all the speed observations per road segment during the
  morning rush hours (8 -- 9 am on weekdays), and compute the standard
  deviation of the speeds, normalized with respect to its mean -- we call this
  the \emph{coefficient of variation}.

  \item We setup a Warp:AdHoc execution environment on two different
  micro-clusters: (1) Cluster 1 with 300 servers with a total equivalent of 965
  cores of Intel Haswell Xeon (2.3GHz) processors and 3.5 TB of RAM, and (2)
  Cluster 2 with 110 servers with a total equivalent of 118 cores and 330 GB of
  RAM. Note that these clusters have about 13\% and 1.2\% RAM respectively
  relative to the dataset size, and cost about 5$\times$ and 40$\times$ lower
  when compared to a cluster with 100\% RAM capacity as required by other
  main-memory based techniques discussed in Section \ref{related_work}.

\end{itemize}

\begin{figure}[h]
	\centering
	\includegraphics[width=3.3in]{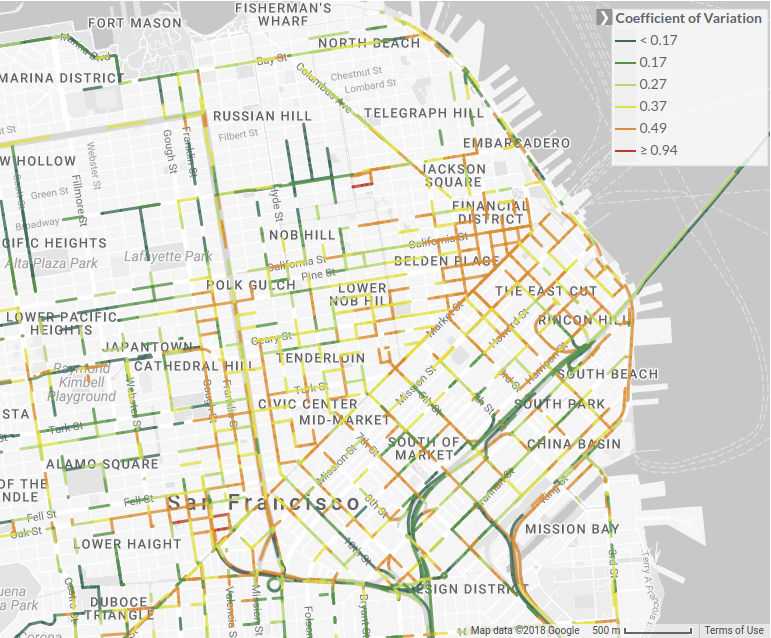}
	\caption{Q1: Normalized traffic speed variation in SF.}
	\label{fig:traffic_variability}
\end{figure}

We run below series of queries on this dataset to compute traffic speed
variations over different geospatial and time regions. Then to visualize the
speed variations on roads for query Q1, we join the resulting data with the
dataset of road geometry and render them on a map. For example, Figure
\ref{fig:traffic_variability} shows the variations for roads in San Francisco.

\begin{itemize}

  \item Q1: San Francisco over a month

  \item Q2: San Francisco over 6 months

  \item Q3: San Francisco Bay Area (including Berkeley, South Bay, Fremont,
  etc.) over a month

  \item Q4: San Francisco Bay Area over 6 months

  \item Q5: California over a month

\end{itemize}

\begin{table}[h]
  \begin{center}
  \begin{tabular}{ l r r }
    \toprule
    Query & CPU time & Exec. time \\

    \midrule
    Geospatial index  & 5.4\,h & 2.2\,m  \\
    Multiple indices  & 45\,m  & 17.6\,s \\
    10\% sample       &  4\,m  & 12.0\,s \\
    1\% sample        & 23\,s  & 11.2\,s \\
    \bottomrule
  \end{tabular}
  \caption{Performance metrics for Q1 on Cluster 1 under different selection
    criteria.}
  \label{tab:example_query}
  \end{center}
\end{table}

%% Previous table including record-metrics.
% \begin{table}[h]
%   \begin{center}
%   \begin{tabular}{ l r r r r }
%     \toprule
%       & \multicolumn{2}{c}{Records} & \multicolumn{2}{c}{Time} \\
%     \cmidrule(lr){2-3} \cmidrule(lr){4-5}
%     \multicolumn{1}{c}{\multirow{-2}{*}[0.5ex]{Query}}
%      & Num. & Size & CPU & Exec. \\

%     \midrule
%     Geospatial index  &  6.21\,M & 39.4\,GiB & 5.4\,h & 2.2\,m \\
%     Multiple indices  &  600\,K & 3.77\,GiB & 45\,m & 17.6\,s \\
%     10\% sample       & 58.0\,K &  372\,MiB &  4\,m & 12.0\,s \\
%     1\% sample        & 5.60\,K & 33.1\,MiB & 23\,s & 11.2\,s \\
%     \bottomrule
%   \end{tabular}
%   \caption{Performance metrics for Q1 on Cluster 1 under different selection
%     criteria.}
%   \label{tab:example_query}
%   \end{center}
% \end{table}

\begin{figure*}[h]
  \centering
  % \begin{subfigure}{0.25\textwidth}
	% 	\centering
	% 	\includegraphics[width=\textwidth]{exp_data}
	% 	\caption{Query data size}
	% 	\label{fig:exp_data}
	% \end{subfigure}
  % \hfill
	\begin{subfigure}{0.32\textwidth}
		\centering
		\includegraphics[width=\textwidth]{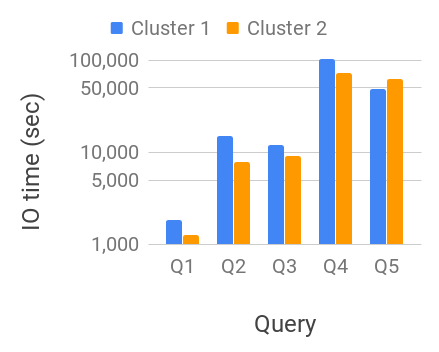}
		\caption{IO time}
		\label{fig:exp_io_time}
	\end{subfigure}
	\hfill
	\begin{subfigure}{0.32\textwidth}
		\centering
		\includegraphics[width=\textwidth]{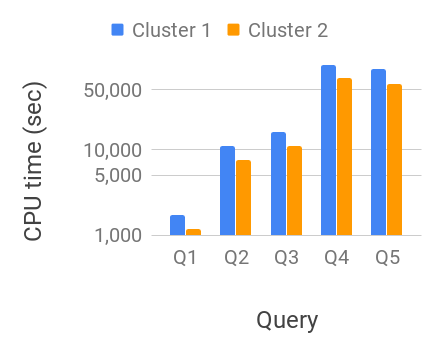}
		\caption{CPU time}
		\label{fig:exp_cpu_time}
	\end{subfigure}
	\hfill
	\begin{subfigure}{0.32\textwidth}
		\centering
		\includegraphics[width=\textwidth]{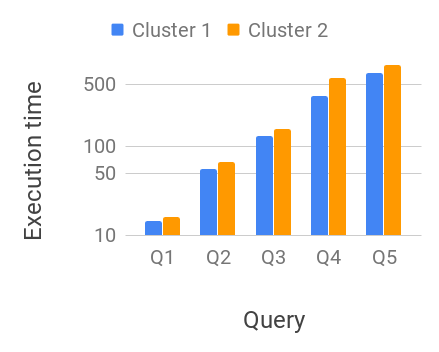}
		\caption{Execution time}
		\label{fig:exp_exec_time}
	\end{subfigure}
	\caption{Performance metrics for different queries on Cluster 1 \& Cluster 2.}
	\label{fig:query_comparison}
\end{figure*}

In addition, we run these queries with different selection criteria, described
below. Table \ref{tab:example_query} shows the total CPU time and the execution
time for Q1 on Cluster 1.

\paragraph*{Geospatial index} Instead of scanning all the records and filtering
them, we use the geospatial indices to only select relevant road segments and
filter out observations that are outside the morning rush hours on weekdays.

\paragraph*{Multiple indices} In addition to the geospatial index, we use
indices on the time of day and day of week to read precisely the data that is
required for the query. This is the most common way of querying with WarpFlow
utilizing the full power of its indices.

\paragraph*{10\% sample} Instead of using the entire dataset, we use a 10\%
sample to get quick estimates at the cost of accuracy (in this case $\sim$5\%
error). Sampling selects only a subset of shards to feed the query. This
results in a slightly faster execution, although not linearly scaled since we
are now using fewer Servers.

\paragraph*{1\% sample} Here we only use a 1\% sample for an even quicker but
cruder estimate (in this case $\sim$20\% error). Although the data scan size
goes down by about a factor of 10, the execution time is very similar to using
a 10\% sample. This is because we gain little from parallelism when using only
1\% of the data shards.

\begin{figure}[h]
	\centering
	\includegraphics[width=3.2in]{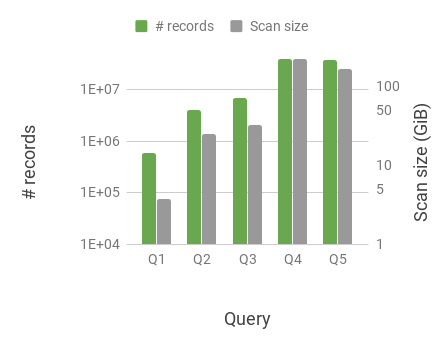}
	\caption{Query data size}
	\label{fig:exp_data}
\end{figure}

Figure \ref{fig:exp_data} shows the data size for different queries. We show
several performance metrics on both the clusters and compare them for these
queries in Figure \ref{fig:query_comparison}. These measurements are averaged
over 5 individual runs. Even though the underlying dataset is much larger
compared to the total available memory, using geospatial and time indices
dramatically reduces the data scan size and consequently IO and CPU costs.
Notice that the number of records relevant to the query increase from Q1
through Q5. The overall data scan size along with IO time, CPU time and the
total execution time scale roughly in the same proportion.

In this performance setup, the total execution times for cluster 2 are only up
to 20\% slower with 8$\times$ reduced CPU capacity and 10$\times$ reduced RAM
capacity, illustrating some of the IO and CPU tradeoffs we discussed in Section
\ref{related_work}. As discussed earlier, we can deploy such micro-clusters in
the cloud in a cost efficient fashion, vs more expensive (5 -- 40$\times$ more)
dedicated clusters necessary for techniques discussed in Section
\ref{related_work}. We also observe some minor variation in IO times (e.g., Q4
vs. Q5), a common occurrence in distributed machine clusters with networked
file systems \cite{misc:aws,misc:gcp}. In addition, the smaller cluster 2 has a
much better efficiency with minimal impact on the performance of the queries.
IO and CPU times are roughly similar when compared with cluster 1. Ideally,
they would have identical IO and CPU times in the absence of any overhead, but
the per-machine overhead slightly increases these times. In fact, they are
somewhat higher for cluster 1 as it has many more machines and hence, a higher
total overhead.

In practice, we see similar behavior in production over \textbf{tens of
thousands} of pipeline runs. Currently, WarpFlow runs on tens of thousand of
cores and about 10 TBs of RAM in a shared cluster, and handles multiple
datasets (tens of terabytes to petascale) stored in a networked file system. We
notice the following workflow working well for developers.

\begin{itemize}

  \item Developers typically begin their analysis with small explorations on
  Warp:AdHoc to get intuition about (say) different cities or small regions,
  and benefit from fast \emph{time-to-first-result}. The interactive execution
  environment on micro-clusters works well because filtered data fits in the
  memory (typically $\sim$ 10s -- 100s of GB) even if the datasets are much
  larger (10s of TB -- PB).

  \item Developers then train these data slices to iterate on learning models
  and get intuition with TensorFlow operators, and get fast
  \emph{time-to-trained-model}.

  \item For further analyses over countries, they use batch execution with
  Warp:Flume that can autoscale the resources, and use both RAM and persistent
  disk storage optimized for a multi-step, reliable execution. By using the
  same WFL query in a batch execution environment, this results in a faster
  \emph{time-to-full-scale-result} as well.

\end{itemize}

\section{Conclusions} \label{conclusions}

WarpFlow is a fast, interactive querying system that speeds up developer
workflows by reducing the time from ideation to prototyping to validation. In
this paper, we focus on Tesseract queries on big and noisy spatiotemporal
datasets. We discuss our indexing structures on Protocol Buffers, FDb,
optimized for fast data selection with extensive indexing and machine learning
support. We presented two execution engines: Warp:AdHoc -- an ad hoc,
interactive version, and Warp:Flume -- a batch processing execution engine
built on Flume. We discussed our developers' experience in running queries in a
cost-efficient manner on micro-clusters of different sizes. WarpFlow's
techniques work well in a shared cluster environment, a practical requirement
for large-scale data development. With WFL and dual-execution modes, WarpFlow's
developers gain a significant speedup in \emph{time-to-first-result},
\emph{time-to-full-scale-result}, and \emph{time-to-trained-model}.

\bibliographystyle{ACM-Reference-Format}
\bibliography{warp_flow}


\begin{thebibliography}{42}

%%% the Web address.

\ifx \showCODEN    \undefined \def \showCODEN     #1{\unskip}     \fi
\ifx \showDOI      \undefined \def \showDOI       #1{#1}\fi
\ifx \showISBNx    \undefined \def \showISBNx     #1{\unskip}     \fi
\ifx \showISBNxiii \undefined \def \showISBNxiii  #1{\unskip}     \fi
\ifx \showISSN     \undefined \def \showISSN      #1{\unskip}     \fi
\ifx \showLCCN     \undefined \def \showLCCN      #1{\unskip}     \fi
\ifx \shownote     \undefined \def \shownote      #1{#1}          \fi
\ifx \showarticletitle \undefined \def \showarticletitle #1{#1}   \fi
\ifx \showURL      \undefined \def \showURL       {\relax}        \fi
% invisible to TeX
\providecommand\bibfield[2]{#2}
\providecommand\bibinfo[2]{#2}
\providecommand\natexlab[1]{#1}
\providecommand\showeprint[2][]{arXiv:#2}

\bibitem[\protect\citeauthoryear{??}{mis}{2011}]%
        {misc:leveldb}
 \bibinfo{year}{2011}\natexlab{}.
\newblock \bibinfo{title}{{Google Open Source Blog: LevelDB: A Fast Persistent
  Key-Value Store}}.
\newblock
\newblock
\urldef\tempurl%
\url{http://google-opensource.blogspot.com/2011/07/leveldb-fast-persistent-key-value-store.html}
\showURL{%
Retrieved March 23, 2018 from \tempurl}


\bibitem[\protect\citeauthoryear{??}{mis}{2014}]%
        {misc:recordio}
 \bibinfo{year}{2014}\natexlab{}.
\newblock \bibinfo{title}{{RecordIO for Google App Engine}}.
\newblock
\newblock
\urldef\tempurl%
\url{https://github.com/n-dream/recordio}
\showURL{%
Retrieved March 23, 2018 from \tempurl}


\bibitem[\protect\citeauthoryear{??}{mis}{2017a}]%
        {misc:gmapskm}
 \bibinfo{year}{2017}\natexlab{a}.
\newblock \bibinfo{title}{{Making AI work for everyone}}.
\newblock
\newblock
\urldef\tempurl%
\url{https://www.blog.google/topics/machine-learning/making-ai-work-for-everyone/}
\showURL{%
Retrieved April 25, 2018 from \tempurl}


\bibitem[\protect\citeauthoryear{??}{mis}{2017b}]%
        {misc:protobuf}
 \bibinfo{year}{2017}\natexlab{b}.
\newblock \bibinfo{title}{{Protocol Buffers: Developer Guide}}.
\newblock
\newblock
\urldef\tempurl%
\url{https://developers.google.com/protocol-buffers/docs/overview}
\showURL{%
Retrieved March 21, 2018 from \tempurl}


\bibitem[\protect\citeauthoryear{??}{mis}{2017c}]%
        {misc:dynproto}
 \bibinfo{year}{2017}\natexlab{c}.
\newblock \bibinfo{title}{{Protocol Buffers: Techniques}}.
\newblock
\newblock
\urldef\tempurl%
\url{https://developers.google.com/protocol-buffers/docs/techniques}
\showURL{%
Retrieved March 21, 2018 from \tempurl}


\bibitem[\protect\citeauthoryear{??}{mis}{2018a}]%
        {misc:aws}
 \bibinfo{year}{2018}\natexlab{a}.
\newblock \bibinfo{title}{{Amazon Web Services}}.
\newblock
\newblock
\urldef\tempurl%
\url{https://aws.amazon.com/}
\showURL{%
Retrieved July 2, 2018 from \tempurl}


\bibitem[\protect\citeauthoryear{??}{mis}{2018b}]%
        {misc:tpu}
 \bibinfo{year}{2018}\natexlab{b}.
\newblock \bibinfo{title}{{Cloud TPU}}.
\newblock
\newblock
\urldef\tempurl%
\url{https://cloud.google.com/tpu/}
\showURL{%
Retrieved March 23, 2018 from \tempurl}


\bibitem[\protect\citeauthoryear{??}{mis}{2018c}]%
        {misc:colab}
 \bibinfo{year}{2018}\natexlab{c}.
\newblock \bibinfo{title}{{Colaboratory}}.
\newblock
\newblock
\urldef\tempurl%
\url{https://research.google.com/colaboratory/faq.html}
\showURL{%
Retrieved April 24, 2018 from \tempurl}


\bibitem[\protect\citeauthoryear{??}{mis}{2018d}]%
        {misc:gcp}
 \bibinfo{year}{2018}\natexlab{d}.
\newblock \bibinfo{title}{{Google Cloud}}.
\newblock
\newblock
\urldef\tempurl%
\url{https://cloud.google.com/}
\showURL{%
Retrieved July 2, 2018 from \tempurl}


\bibitem[\protect\citeauthoryear{??}{mis}{2018e}]%
        {misc:gmapsusers}
 \bibinfo{year}{2018}\natexlab{e}.
\newblock \bibinfo{title}{{Google Maps learns 39 new languages}}.
\newblock
\newblock
\urldef\tempurl%
\url{https://www.blog.google/products/maps/google-maps-learns-39-new-languages/}
\showURL{%
Retrieved April 25, 2018 from \tempurl}


\bibitem[\protect\citeauthoryear{??}{mis}{2018f}]%
        {misc:azure}
 \bibinfo{year}{2018}\natexlab{f}.
\newblock \bibinfo{title}{{Microsoft Azure Cloud Computing Platform \&
  Services}}.
\newblock
\newblock
\urldef\tempurl%
\url{https://azure.microsoft.com/en-us/}
\showURL{%
Retrieved July 2, 2018 from \tempurl}


\bibitem[\protect\citeauthoryear{??}{mis}{2018g}]%
        {misc:mysql}
 \bibinfo{year}{2018}\natexlab{g}.
\newblock \bibinfo{title}{{MySQL}}.
\newblock
\newblock
\urldef\tempurl%
\url{https://www.mysql.com/}
\showURL{%
Retrieved July 2, 2018 from \tempurl}


\bibitem[\protect\citeauthoryear{??}{mis}{2018h}]%
        {misc:postgis}
 \bibinfo{year}{2018}\natexlab{h}.
\newblock \bibinfo{title}{{PostGIS}}.
\newblock
\newblock
\urldef\tempurl%
\url{https://postgis.net}
\showURL{%
Retrieved March 28, 2018 from \tempurl}


\bibitem[\protect\citeauthoryear{??}{mis}{2018i}]%
        {misc:postgres}
 \bibinfo{year}{2018}\natexlab{i}.
\newblock \bibinfo{title}{{PostgreSQL}}.
\newblock
\newblock
\urldef\tempurl%
\url{https://www.postgresql.org}
\showURL{%
Retrieved March 28, 2018 from \tempurl}


\bibitem[\protect\citeauthoryear{??}{mis}{2018j}]%
        {misc:repl}
 \bibinfo{year}{2018}\natexlab{j}.
\newblock \bibinfo{title}{{Scala REPL: Overview}}.
\newblock
\newblock
\urldef\tempurl%
\url{https://docs.scala-lang.org/overviews/repl/overview.html}
\showURL{%
Retrieved March 23, 2018 from \tempurl}


\bibitem[\protect\citeauthoryear{??}{mis}{2018k}]%
        {misc:tensorflow}
 \bibinfo{year}{2018}\natexlab{k}.
\newblock \bibinfo{title}{{TensorFlow: An open-source machine learning
  framework for everyone}}.
\newblock
\newblock
\urldef\tempurl%
\url{https://www.tensorflow.org/}
\showURL{%
Retrieved March 23, 2018 from \tempurl}


\bibitem[\protect\citeauthoryear{??}{mis}{2018l}]%
        {misc:savedmodel}
 \bibinfo{year}{2018}\natexlab{l}.
\newblock \bibinfo{title}{{TensorFlow: Saving and Restoring}}.
\newblock
\newblock
\urldef\tempurl%
\url{https://www.tensorflow.org/programmers_guide/saved_model}
\showURL{%
Retrieved March 23, 2018 from \tempurl}


\bibitem[\protect\citeauthoryear{??}{mis}{2018m}]%
        {misc:jupyter}
 \bibinfo{year}{2018}\natexlab{m}.
\newblock \bibinfo{title}{{The Jupyter Notebook}}.
\newblock
\newblock
\urldef\tempurl%
\url{https://jupyter.org/}
\showURL{%
Retrieved April 24, 2018 from \tempurl}


\bibitem[\protect\citeauthoryear{??}{mis}{2018n}]%
        {misc:scala}
 \bibinfo{year}{2018}\natexlab{n}.
\newblock \bibinfo{title}{{The Scala Programming Language}}.
\newblock
\newblock
\urldef\tempurl%
\url{https://www.scala-lang.org/}
\showURL{%
Retrieved April 16, 2018 from \tempurl}


\bibitem[\protect\citeauthoryear{Bloom}{Bloom}{1970}]%
        {bloom:filters}
\bibfield{author}{\bibinfo{person}{Burton~H. Bloom}.}
  \bibinfo{year}{1970}\natexlab{}.
\newblock \showarticletitle{{Space/Time Trade-offs in Hash Coding with
  Allowable Errors}}.
\newblock \bibinfo{journal}{\emph{Commun. ACM}} \bibinfo{volume}{13},
  \bibinfo{number}{7} (\bibinfo{date}{July} \bibinfo{year}{1970}),
  \bibinfo{pages}{422--426}.
\newblock
\showISSN{0001-0782}
\urldef\tempurl%
\url{https://doi.org/10.1145/362686.362692}
\showDOI{\tempurl}


\bibitem[\protect\citeauthoryear{Chang, Dean, Ghemawat, Hsieh, Wallach,
  Burrows, Chandra, Fikes, and Gruber}{Chang et~al\mbox{.}}{2008}]%
        {chang:bigtable}
\bibfield{author}{\bibinfo{person}{Fay Chang}, \bibinfo{person}{Jeffrey Dean},
  \bibinfo{person}{Sanjay Ghemawat}, \bibinfo{person}{Wilson~C. Hsieh},
  \bibinfo{person}{Deborah~A. Wallach}, \bibinfo{person}{Mike Burrows},
  \bibinfo{person}{Tushar Chandra}, \bibinfo{person}{Andrew Fikes}, {and}
  \bibinfo{person}{Robert~E. Gruber}.} \bibinfo{year}{2008}\natexlab{}.
\newblock \showarticletitle{{Bigtable: A Distributed Storage System for
  Structured Data}}.
\newblock \bibinfo{journal}{\emph{ACM Trans. Comput. Syst.}}
  \bibinfo{volume}{26}, \bibinfo{number}{2}, Article \bibinfo{articleno}{4}
  (\bibinfo{date}{June} \bibinfo{year}{2008}), \bibinfo{numpages}{26}~pages.
\newblock
\showISSN{0734-2071}
\urldef\tempurl%
\url{https://doi.org/10.1145/1365815.1365816}
\showDOI{\tempurl}


\bibitem[\protect\citeauthoryear{Cormen, Leiserson, Rivest, and Stein}{Cormen
  et~al\mbox{.}}{2009}]%
        {cormen:algo}
\bibfield{author}{\bibinfo{person}{Thomas~H. Cormen},
  \bibinfo{person}{Charles~E. Leiserson}, \bibinfo{person}{Ronald~L. Rivest},
  {and} \bibinfo{person}{Clifford Stein}.} \bibinfo{year}{2009}\natexlab{}.
\newblock \bibinfo{booktitle}{\emph{{Introduction to Algorithms, Third
  Edition}} (\bibinfo{edition}{3rd} ed.)}.
\newblock \bibinfo{publisher}{The MIT Press}, \bibinfo{pages}{348--356}.
\newblock
\showISBNx{0262033844, 9780262033848}


\bibitem[\protect\citeauthoryear{Costa, Chatzimilioudis, Zeinalipour-Yazti, and
  Mokbel}{Costa et~al\mbox{.}}{2017}]%
        {costa:spate}
\bibfield{author}{\bibinfo{person}{C. Costa}, \bibinfo{person}{G.
  Chatzimilioudis}, \bibinfo{person}{D. Zeinalipour-Yazti}, {and}
  \bibinfo{person}{M.~F. Mokbel}.} \bibinfo{year}{2017}\natexlab{}.
\newblock \showarticletitle{{Efficient Exploration of Telco Big Data with
  Compression and Decaying}}. In \bibinfo{booktitle}{\emph{2017 IEEE 33rd
  International Conference on Data Engineering (ICDE)}}.
  \bibinfo{pages}{1332--1343}.
\newblock
\urldef\tempurl%
\url{https://doi.org/10.1109/ICDE.2017.175}
\showDOI{\tempurl}


\bibitem[\protect\citeauthoryear{Dean and Ghemawat}{Dean and Ghemawat}{2008}]%
        {dean:mr}
\bibfield{author}{\bibinfo{person}{Jeffrey Dean} {and} \bibinfo{person}{Sanjay
  Ghemawat}.} \bibinfo{year}{2008}\natexlab{}.
\newblock \showarticletitle{{MapReduce: Simplified Data Processing on Large
  Clusters}}.
\newblock \bibinfo{journal}{\emph{Commun. ACM}} \bibinfo{volume}{51},
  \bibinfo{number}{1} (\bibinfo{date}{Jan.} \bibinfo{year}{2008}),
  \bibinfo{pages}{107--113}.
\newblock
\showISSN{0001-0782}
\urldef\tempurl%
\url{https://doi.org/10.1145/1327452.1327492}
\showDOI{\tempurl}


\bibitem[\protect\citeauthoryear{Eldawy}{Eldawy}{2014}]%
        {eldawy:spatialhadoop}
\bibfield{author}{\bibinfo{person}{Ahmed Eldawy}.}
  \bibinfo{year}{2014}\natexlab{}.
\newblock \showarticletitle{{SpatialHadoop: Towards Flexible and Scalable
  Spatial Processing Using Mapreduce}}. In
  \bibinfo{booktitle}{\emph{Proceedings of the 2014 SIGMOD PhD Symposium}}
  \emph{(\bibinfo{series}{SIGMOD'14 PhD Symposium})}. \bibinfo{publisher}{ACM},
  \bibinfo{address}{New York, NY, USA}, \bibinfo{pages}{46--50}.
\newblock
\showISBNx{978-1-4503-2924-8}
\urldef\tempurl%
\url{https://doi.org/10.1145/2602622.2602625}
\showDOI{\tempurl}


\bibitem[\protect\citeauthoryear{et~al.}{et~al.}{2010}]%
        {chambers:flume}
\bibfield{author}{\bibinfo{person}{Craig~Chambers et al.}}
  \bibinfo{year}{2010}\natexlab{}.
\newblock \showarticletitle{{FlumeJava: Easy, Efficient Data-parallel
  Pipelines}}. In \bibinfo{booktitle}{\emph{Proceedings of the 31st ACM SIGPLAN
  Conference on Programming Language Design and Implementation}}
  \emph{(\bibinfo{series}{PLDI '10})}. \bibinfo{publisher}{ACM},
  \bibinfo{address}{New York, NY, USA}, \bibinfo{pages}{363--375}.
\newblock
\showISBNx{978-1-4503-0019-3}
\urldef\tempurl%
\url{https://doi.org/10.1145/1806596.1806638}
\showDOI{\tempurl}


\bibitem[\protect\citeauthoryear{et~al.}{et~al.}{2016}]%
        {manoharan:shasta}
\bibfield{author}{\bibinfo{person}{Gokul Nath Babu~Manoharan et al.}}
  \bibinfo{year}{2016}\natexlab{}.
\newblock \showarticletitle{{Shasta: Interactive Reporting At Scale}}. In
  \bibinfo{booktitle}{\emph{Proceedings of the 2016 International Conference on
  Management of Data}} \emph{(\bibinfo{series}{SIGMOD '16})}.
  \bibinfo{publisher}{ACM}, \bibinfo{address}{New York, NY, USA},
  \bibinfo{pages}{1393--1404}.
\newblock
\showISBNx{978-1-4503-3531-7}
\urldef\tempurl%
\url{https://doi.org/10.1145/2882903.2904444}
\showDOI{\tempurl}


\bibitem[\protect\citeauthoryear{et~al.}{et~al.}{2013}]%
        {shute:f1}
\bibfield{author}{\bibinfo{person}{Jeff~Shute et al.}}
  \bibinfo{year}{2013}\natexlab{}.
\newblock \showarticletitle{{F1: A Distributed SQL Database That Scales}}.
\newblock \bibinfo{journal}{\emph{Proc. VLDB Endow.}} \bibinfo{volume}{6},
  \bibinfo{number}{11} (\bibinfo{date}{Aug.} \bibinfo{year}{2013}),
  \bibinfo{pages}{1068--1079}.
\newblock
\showISSN{2150-8097}
\urldef\tempurl%
\url{https://doi.org/10.14778/2536222.2536232}
\showDOI{\tempurl}


\bibitem[\protect\citeauthoryear{Finkel and Bentley}{Finkel and
  Bentley}{1974}]%
        {finkel:quadtree}
\bibfield{author}{\bibinfo{person}{R.~A. Finkel} {and} \bibinfo{person}{J.~L.
  Bentley}.} \bibinfo{year}{1974}\natexlab{}.
\newblock \showarticletitle{{Quad Trees a Data Structure for Retrieval on
  Composite Keys}}.
\newblock \bibinfo{journal}{\emph{Acta Inf.}} \bibinfo{volume}{4},
  \bibinfo{number}{1} (\bibinfo{date}{March} \bibinfo{year}{1974}),
  \bibinfo{pages}{1--9}.
\newblock
\showISSN{0001-5903}
\urldef\tempurl%
\url{https://doi.org/10.1007/BF00288933}
\showDOI{\tempurl}


\bibitem[\protect\citeauthoryear{Flajolet, Fusy, Gandouet, and
  Meunier}{Flajolet et~al\mbox{.}}{2007}]%
        {flajolet:hll}
\bibfield{author}{\bibinfo{person}{Philippe Flajolet},
  \bibinfo{person}{{\'E}ric Fusy}, \bibinfo{person}{Olivier Gandouet}, {and}
  \bibinfo{person}{Fr{\'e}d{\'e}ric Meunier}.} \bibinfo{year}{2007}\natexlab{}.
\newblock \showarticletitle{{HyperLogLog: the analysis of a near-optimal
  cardinality estimation algorithm}}. In \bibinfo{booktitle}{\emph{{AofA:
  Analysis of Algorithms}}} \emph{(\bibinfo{series}{DMTCS Proceedings})},
  \bibfield{editor}{\bibinfo{person}{Philippe Jacquet}} (Ed.),
  Vol.~\bibinfo{volume}{DMTCS Proceedings vol. AH, 2007}.
  \bibinfo{publisher}{{Discrete Mathematics and Theoretical Computer Science}},
  \bibinfo{address}{Juan les Pins, France}, \bibinfo{pages}{137--156}.
\newblock


\bibitem[\protect\citeauthoryear{Garcia-Molina, Ullman, and
  Widom}{Garcia-Molina et~al\mbox{.}}{2008}]%
        {hector:database}
\bibfield{author}{\bibinfo{person}{Hector Garcia-Molina},
  \bibinfo{person}{Jeffrey~D. Ullman}, {and} \bibinfo{person}{Jennifer Widom}.}
  \bibinfo{year}{2008}\natexlab{}.
\newblock \bibinfo{booktitle}{\emph{{Database Systems: The Complete Book}}
  (\bibinfo{edition}{second} ed.)}.
\newblock \bibinfo{publisher}{Pearson Prentice Hall}, \bibinfo{address}{Upper
  Saddle River, NJ, USA}, \bibinfo{pages}{728,734,742,743,824,829}.
\newblock


\bibitem[\protect\citeauthoryear{Lee, Jones, Ridenour, Bennett, Majors, Melito,
  and Wilson}{Lee et~al\mbox{.}}{2016}]%
        {lee:gps}
\bibfield{author}{\bibinfo{person}{L. Lee}, \bibinfo{person}{M. Jones},
  \bibinfo{person}{G.~S. Ridenour}, \bibinfo{person}{S.~J. Bennett},
  \bibinfo{person}{A.~C. Majors}, \bibinfo{person}{B.~L. Melito}, {and}
  \bibinfo{person}{M.~J. Wilson}.} \bibinfo{year}{2016}\natexlab{}.
\newblock \showarticletitle{{Comparison of Accuracy and Precision of
  GPS-Enabled Mobile Devices}}. In \bibinfo{booktitle}{\emph{2016 IEEE
  International Conference on Computer and Information Technology (CIT)}}.
  \bibinfo{pages}{73--82}.
\newblock
\urldef\tempurl%
\url{https://doi.org/10.1109/CIT.2016.94}
\showDOI{\tempurl}


\bibitem[\protect\citeauthoryear{Maling}{Maling}{2013}]%
        {maling:coord}
\bibfield{author}{\bibinfo{person}{D.H. Maling}.}
  \bibinfo{year}{2013}\natexlab{}.
\newblock \bibinfo{booktitle}{\emph{{Coordinate System and Map Projections}}
  (\bibinfo{edition}{second} ed.)}.
\newblock \bibinfo{publisher}{Pergamon Press}, \bibinfo{address}{Elmsford, NY,
  USA}, \bibinfo{pages}{336--363}.
\newblock


\bibitem[\protect\citeauthoryear{Melnik, Gubarev, Long, Romer, Shivakumar,
  Tolton, and Vassilakis}{Melnik et~al\mbox{.}}{2010}]%
        {melnik:dremel}
\bibfield{author}{\bibinfo{person}{Sergey Melnik}, \bibinfo{person}{Andrey
  Gubarev}, \bibinfo{person}{Jing~Jing Long}, \bibinfo{person}{Geoffrey Romer},
  \bibinfo{person}{Shiva Shivakumar}, \bibinfo{person}{Matt Tolton}, {and}
  \bibinfo{person}{Theo Vassilakis}.} \bibinfo{year}{2010}\natexlab{}.
\newblock \showarticletitle{{Dremel: Interactive Analysis of Web-scale
  Datasets}}.
\newblock \bibinfo{journal}{\emph{Proc. VLDB Endow.}} \bibinfo{volume}{3},
  \bibinfo{number}{1-2}, \bibinfo{pages}{330--339}.
\newblock
\showISSN{2150-8097}
\urldef\tempurl%
\url{https://doi.org/10.14778/1920841.1920886}
\showDOI{\tempurl}


\bibitem[\protect\citeauthoryear{Palkar, Thomas, Shanbhag, Narayanan, Pirk,
  Schwarzkopf, Amarasinghe, Zaharia, and InfoLab}{Palkar et~al\mbox{.}}{2016}]%
        {shoumik:weld}
\bibfield{author}{\bibinfo{person}{Shoumik Palkar}, \bibinfo{person}{J.~Joshua
  Thomas}, \bibinfo{person}{Anil Shanbhag}, \bibinfo{person}{Deepak Narayanan},
  \bibinfo{person}{Holger Pirk}, \bibinfo{person}{Malte Schwarzkopf},
  \bibinfo{person}{Saman Amarasinghe}, \bibinfo{person}{Matei Zaharia}, {and}
  \bibinfo{person}{Stanford InfoLab}.} \bibinfo{year}{2016}\natexlab{}.
\newblock \showarticletitle{{Weld : A Common Runtime for High Performance Data
  Analytics}}.
\newblock


\bibitem[\protect\citeauthoryear{Shang, Li, and Bao}{Shang
  et~al\mbox{.}}{2018}]%
        {shang:dita}
\bibfield{author}{\bibinfo{person}{Zeyuan Shang}, \bibinfo{person}{Guoliang
  Li}, {and} \bibinfo{person}{Zhifeng Bao}.} \bibinfo{year}{2018}\natexlab{}.
\newblock \showarticletitle{{DITA: Distributed In-Memory Trajectory
  Analytics}}. In \bibinfo{booktitle}{\emph{Proceedings of the 2018
  International Conference on Management of Data}}
  \emph{(\bibinfo{series}{SIGMOD '18})}. \bibinfo{publisher}{ACM},
  \bibinfo{address}{New York, NY, USA}, \bibinfo{pages}{725--740}.
\newblock
\showISBNx{978-1-4503-4703-7}
\urldef\tempurl%
\url{https://doi.org/10.1145/3183713.3183743}
\showDOI{\tempurl}


\bibitem[\protect\citeauthoryear{von Watzdorf and Michahelles}{von Watzdorf and
  Michahelles}{2010}]%
        {vonWatzdorf:gps}
\bibfield{author}{\bibinfo{person}{Stephan von Watzdorf} {and}
  \bibinfo{person}{Florian Michahelles}.} \bibinfo{year}{2010}\natexlab{}.
\newblock \showarticletitle{{Accuracy of Positioning Data on Smartphones}}. In
  \bibinfo{booktitle}{\emph{Proceedings of the 3rd International Workshop on
  Location and the Web}} \emph{(\bibinfo{series}{LocWeb '10})}.
  \bibinfo{publisher}{ACM}, \bibinfo{address}{New York, NY, USA}, Article
  \bibinfo{articleno}{2}, \bibinfo{numpages}{4}~pages.
\newblock
\showISBNx{978-1-4503-0412-2}
\urldef\tempurl%
\url{https://doi.org/10.1145/1899662.1899664}
\showDOI{\tempurl}


\bibitem[\protect\citeauthoryear{Xie, Li, and Phillips}{Xie
  et~al\mbox{.}}{2017}]%
        {xie:simbatraj}
\bibfield{author}{\bibinfo{person}{Dong Xie}, \bibinfo{person}{Feifei Li},
  {and} \bibinfo{person}{Jeff~M. Phillips}.} \bibinfo{year}{2017}\natexlab{}.
\newblock \showarticletitle{{Distributed Trajectory Similarity Search}}.
\newblock \bibinfo{journal}{\emph{Proc. VLDB Endow.}} \bibinfo{volume}{10},
  \bibinfo{number}{11} (\bibinfo{date}{Aug.} \bibinfo{year}{2017}),
  \bibinfo{pages}{1478--1489}.
\newblock
\showISSN{2150-8097}
\urldef\tempurl%
\url{https://doi.org/10.14778/3137628.3137655}
\showDOI{\tempurl}


\bibitem[\protect\citeauthoryear{Xie, Li, Yao, Li, Zhou, and Guo}{Xie
  et~al\mbox{.}}{2016}]%
        {xie:simba}
\bibfield{author}{\bibinfo{person}{Dong Xie}, \bibinfo{person}{Feifei Li},
  \bibinfo{person}{Bin Yao}, \bibinfo{person}{Gefei Li}, \bibinfo{person}{Liang
  Zhou}, {and} \bibinfo{person}{Minyi Guo}.} \bibinfo{year}{2016}\natexlab{}.
\newblock \showarticletitle{{Simba: Efficient In-Memory Spatial Analytics}}. In
  \bibinfo{booktitle}{\emph{Proceedings of the 2016 International Conference on
  Management of Data}} \emph{(\bibinfo{series}{SIGMOD '16})}.
  \bibinfo{publisher}{ACM}, \bibinfo{address}{New York, NY, USA},
  \bibinfo{pages}{1071--1085}.
\newblock
\showISBNx{978-1-4503-3531-7}
\urldef\tempurl%
\url{https://doi.org/10.1145/2882903.2915237}
\showDOI{\tempurl}


\bibitem[\protect\citeauthoryear{Yu, Wu, and Sarwat}{Yu et~al\mbox{.}}{2015}]%
        {yu:geospark}
\bibfield{author}{\bibinfo{person}{Jia Yu}, \bibinfo{person}{Jinxuan Wu}, {and}
  \bibinfo{person}{Mohamed Sarwat}.} \bibinfo{year}{2015}\natexlab{}.
\newblock \showarticletitle{{GeoSpark: A Cluster Computing Framework for
  Processing Large-scale Spatial Data}}. In
  \bibinfo{booktitle}{\emph{Proceedings of the 23rd SIGSPATIAL International
  Conference on Advances in Geographic Information Systems}}
  \emph{(\bibinfo{series}{SIGSPATIAL '15})}. \bibinfo{publisher}{ACM},
  \bibinfo{address}{New York, NY, USA}, Article \bibinfo{articleno}{70},
  \bibinfo{numpages}{4}~pages.
\newblock
\showISBNx{978-1-4503-3967-4}
\urldef\tempurl%
\url{https://doi.org/10.1145/2820783.2820860}
\showDOI{\tempurl}


\bibitem[\protect\citeauthoryear{Zaharia, Chowdhury, Franklin, Shenker, and
  Stoica}{Zaharia et~al\mbox{.}}{2010}]%
        {zaharia:spark}
\bibfield{author}{\bibinfo{person}{Matei Zaharia}, \bibinfo{person}{Mosharaf
  Chowdhury}, \bibinfo{person}{Michael~J. Franklin}, \bibinfo{person}{Scott
  Shenker}, {and} \bibinfo{person}{Ion Stoica}.}
  \bibinfo{year}{2010}\natexlab{}.
\newblock \showarticletitle{{Spark: Cluster Computing with Working Sets}}. In
  \bibinfo{booktitle}{\emph{Proceedings of the 2nd USENIX Conference on Hot
  Topics in Cloud Computing}} \emph{(\bibinfo{series}{HotCloud'10})}.
  \bibinfo{publisher}{USENIX Association}, \bibinfo{address}{Berkeley, CA,
  USA}, \bibinfo{pages}{10}.
\newblock
\urldef\tempurl%
\url{http://dl.acm.org/citation.cfm?id=1863103.1863113}
\showURL{%
\tempurl}


\bibitem[\protect\citeauthoryear{Zandbergen and Barbeau}{Zandbergen and
  Barbeau}{2011}]%
        {zandbergen:gps}
\bibfield{author}{\bibinfo{person}{Paul~A. Zandbergen} {and}
  \bibinfo{person}{Sean~J. Barbeau}.} \bibinfo{year}{2011}\natexlab{}.
\newblock \showarticletitle{{Positional Accuracy of Assisted GPS Data from
  High-Sensitivity GPS-enabled Mobile Phones}}.
\newblock \bibinfo{journal}{\emph{Journal of Navigation}} \bibinfo{volume}{64},
  \bibinfo{number}{3} (\bibinfo{year}{2011}), \bibinfo{pages}{381–399}.
\newblock
\urldef\tempurl%
\url{https://doi.org/10.1017/S0373463311000051}
\showDOI{\tempurl}


\end{thebibliography}

\end{document}